\documentclass[a4paper, amsfonts, amssymb, amsmath, reprint, showkeys, nofootinbib, twoside]{revtex4-1}
\usepackage[english]{babel}
\usepackage[utf8]{inputenc}
\usepackage[colorinlistoftodos, color=green!40, prependcaption]{todonotes}
\usepackage{amsthm}
\usepackage{mathtools}
\usepackage{physics}
\usepackage{xcolor}
\usepackage{graphicx}
\usepackage[left=23mm,right=13mm,top=35mm,columnsep=15pt]{geometry} 
\usepackage{adjustbox}
\usepackage{placeins}
\usepackage[T1]{fontenc}
\usepackage{lipsum}
\usepackage{csquotes}
\usepackage[pdftex, pdftitle={Article}, pdfauthor={Author}]{hyperref} % For hyperlinks in the PDF
\begin{document}
\title{Study of surface roughness modulation on the adhesion behavior of PDMS elastomer}

\author{Susheel Kumar,\textit{$^{a}$} Krishnacharya Khare,\textit{$^{b}$} Manjesh K. Singh\textit{$^{a}$}}
    \email[Correspondence email address: ]{manjesh@iitk.ac.in}
    \affiliation{$^{a}$~Department of Mechanical Engineering, Indian Institute of Technology Kanpur, Kanpur, India.\\
$^{b}$~Department of Physics, Indian Institute of Technology Kanpur, Kanpur, India.}

% \date{\today} % Leave empty to omit a date

\begin{abstract}
Adhesion control at the interface of two surfaces is crucial in various applications, including the design of micro- and nanodevices such as microfluidic devices, triboelectric nanogenerators, biochips, and electronic sensors. Several factors influence adhesion, including sample preparation, surface energy, mechanical properties such as modulus of elasticity, and surface texture or roughness. This study specifically investigates the effect of surface roughness on the adhesion behavior of the elastomer polydimethylsiloxane (PDMS), focusing on a complementary interface with identical roughness. To create surface roughness, sandpapers with grit sizes ranging from $N = 120$ to $N = 2000$ were used during the molding process. The surface roughness of the PDMS elastomer was then characterized using a stylus profilometer. Interfacial adhesion was evaluated through wedge test experiments, which enabled the analysis of the relationship between surface roughness, work of adhesion, and equilibrium crack length. Furthermore, the study also explores the correlation between the real area of contact and the work of adhesion.
\end{abstract}


\maketitle
\section{Introduction}

The interface between two contacting bodies is considered adhesive when a finite amount of force is required to separate them, indicating the presence of interfacial attractive forces. These forces predominantly arise from van der Waals interactions at the micro- and nanoscale. Such adhesive interactions are fundamental to a variety of real-world and engineering applications, including the remarkable climbing ability of geckos on vertical surfaces \cite{niewiarowski2016sticking}, the human finger’s capacity to grip and manipulate objects \cite{ayyildiz2018contact}, and the development of engineered adhesive mimics inspired by biological systems \cite{geim2003microfabricated}. In parallel, adhesive phenomena are critically leveraged in pick-and-place processes, which are widely utilized across different scales in manufacturing. These range from large-scale industrial assembly lines \cite{popa2004micro} to highly precise nanoscale transfer printing techniques \cite{carlson2012transfer}. By enabling accurate positioning and integration of components, these methods play an essential role in diverse applications spanning electronics fabrication to micro- and nanotechnology. \par
Nature shows many examples where insects have internal parts that attach to each other by fitting together precisely. The shapes of these parts are made to match each other perfectly so they can connect securely. A well-known example is the invention of Velcro, which is based on a hook-loop adhesion mechanism \cite{ouyang2022mechanical}. At much smaller scales, nature employs efficient attachment systems, such as wing-locking mechanisms that hold insects' wings securely in place when they are not in flight \cite{sherge2001biological}. Additionally, some dragonflies possess specialized structures in their necks that help stabilize their heads \cite{gorb1999evolution, sherge2001biological}. These practical applications require the controlled modulation of adhesive forces to achieve sufficient grip and stability during operation while also enabling efficient and timely detachment when required. \par

In particular, we have investigated the adhesion behavior of an elastomeric polymer, namely polydimethylsiloxane (PDMS), which consists of repeating units of dimethylsiloxane monomers. In contemporary engineering and biomedical applications, polymers are increasingly being utilized as substitutes for metals due to their exceptional properties, including biocompatibility, low weight, durability, widespread availability, and cost-effectiveness. Cross-linking and structural heterogeneity play a crucial role in determining a polymer's thermal~\cite{Manoj_macrolett,MukherjiPRM21}, mechanical~\cite{ain2024insights,ain2024109702,ain202433817,ain2024112955,maurya2022computational,maurya2023computational}, and tribological~\cite{singh2016effect,singh2018combined}. properties, thereby enhancing its suitability for targeted applications. \par
The distinctive structural characteristics of PDMS elastomer have attracted considerable attention across interdisciplinary fields of surface science due to its
chemical and thermal stability, hydrophobicity, low surface energy, flexibility, and remarkable capability to accurately replicate surface features \cite{packham2003surface}. These exceptional properties of PDMS elastomer make it widely used in applications such as biomedical devices, soft lithography, sealants, and microfluidics. In these applications of PDMS elastomer, interfacial adhesion between contacting surfaces is crucial and is influenced by various factors, including surface chemistry, the mechanical properties of the material, and the surface topography or roughness at the interface. PDMS elastomer possesses inherently low surface energy, which presents challenges in achieving effective bonding with contacting surfaces. This limitation has highlighted the need to modify the mechanical properties and implement topographical and chemical surface modifications to enhance the adhesion performance of PDMS elastomer. \par
Numerous studies have been conducted to investigate and optimize interfacial adhesion by systematically modifying these factors. Examples of approaches to enhance the adhesion behavior of PDMS elastomers include the use of adhesive interfacial layers, such as GE SS4120 and $\mathrm{Sylgard^{TM}184}$ PDMS \cite{kersey2009effect,tsai2011bonding,samel2007fabrication}, surface modification via oxygen plasma treatment \cite{xiong2014adhesion,bhattacharya2005studies}, chemical treatments such as exposure to trichlorosilane vapor \cite{al2023surface,sofla2010vapor}, and polymer grafting onto the PDMS surface \cite{beh2012pdms}. Furthermore, adhesion can be tuned by modifying the mechanical properties of the PDMS elastomer, such as its modulus of elasticity. This can be achieved by altering the mixing ratio and processing parameters (curing temperature and time) \cite{kumar2025adhesion, kumar2025controlling}. \par

Various techniques are available to evaluate the interfacial adhesion between two contacting surfaces, several of which are outlined below. Adhesion between complementary and non-complementary rippled surfaces of PDMS elastomer has been assessed using the wedge test method \cite{vajpayee2011adhesion}. The results demonstrated that interfacial adhesion increases with greater rippled amplitude, whereas negligible adhesion was observed when the rippled amplitudes were dissimilar. In a related study, a custom-built adhesion tester was employed to evaluate the effect of repeated pull-off forces on the adhesion behavior of PDMS elastomer against a borosilicate glass probe \cite{kroner2010adhesion}. In addition, Atomic Force Microscopy (AFM), specifically using PeakForce Quantitative Nanomechanical Mapping (PF-QNM), was utilized to measure the nanoscale adhesion behavior of PDMS elastomer \cite{nair2019afm}. The peel test was also employed to investigate adhesion characteristics, where the presence of a thin, constrained elastic film led to the formation of instability patterns, the nature of which was influenced by the geometric configuration of the experiment \cite{ghatak2003adhesion}. Further, the spherical probe test was used to examine the adhesion response of PDMS elastomer by varying the mixing ratios of the base and curing agent, highlighting the influence of formulation on adhesive performance \cite{darby2022modulus}. The probe-tack test was conducted to measure the bonding energy of PDMS elastomer, with variations in mechanical properties such as elastic modulus, debonding velocity, and specimen thickness \cite{nase2013debonding}. To study adhesion on different substrates, the 90-degree peel-off test was performed to compare the interfacial adhesion of PDMS elastomer against aluminum and glass substrates \cite{murphy2020tailoring}. Finally, the wedge test approach was also applied to investigate the effects of nanoscale heterogeneity within the bulk material, as well as changes in curing concentration, temperature, and time, on the interfacial adhesion properties of the PDMS elastomer \cite{kumar2025adhesion, kumar2025controlling, r41_Chiranjit}. \par

This study aims to investigate the influence of surface roughness on the adhesion behavior of PDMS elastomer. All surfaces, whether natural or engineered, inherently exhibit a certain degree of roughness, which is crucial in determining interfacial adhesion \cite{luan2005breakdown,jacobs2013effect,pastewka2014contact}. Surface roughness influences the real area of contact, facilitates mechanical interlocking, and modifies the distribution of adhesive forces at the interface, with these effects being particularly relevant in various practical applications. The increased real area of contact between the triboelectric layers, resulting from the textured surface of the PDMS elastomer, enhances the efficiency of charge generation in triboelectric nanogenerators (TENGs). This textured PDMS elastomer is specifically employed to harvest mechanical energy produced by joint movements of the human body, with the harvested energy being used to power wearable electronics and self-powered sensors. In this regard, sandpaper is used to create textured microstructures on the opposing surfaces of the PDMS elastomer and copper electrode, forming a complementary pair that significantly boosts the efficiency of the TENG device \cite{zhang2018high}. Additionally, an aluminum coating technique is employed to fabricate the electrode of the TENG, utilizing sandpaper as the substrate in combination with the textured PDMS elastomer surface to further enhance device performance \cite{kim2017large}.
The triboelectric positivity of the PDMS elastomer is lower than that of human skin, and its high mechanical flexibility further enhances its ability to facilitate efficient charge transfer when in contact with the skin. These characteristics make the PDMS elastomer an attractive candidate for the development of low-cost, polymer-based triboelectric energy harvesting devices, such as self-powered pedometers \cite{rasel2017sandpaper}. Moreover, biomimicry, inspired by the diverse structural and functional characteristics of biological skin, serves as a widely adopted framework for designing synthetic analogs. In this context, the self-healing properties and biomimetic features of PDMS elastomer emphasize its potential for a broad range of bioengineering applications \cite{liu2019design}. In the biomedical field, devices used for vital sign monitoring or drug delivery must adhere securely to the skin, detach without discomfort, and support reuse or safe disposal \cite{yu2018tunable}. Similarly, in microfluidic applications, the structured surface of PDMS elastomer is employed to regulate fluid flow by enhancing capillary effects and minimizing fluidic resistance. These properties are especially advantageous in applications requiring precise control and modulation of fluid flow. To achieve this, surface micromachining techniques, particularly reactive ion etching (RIE), are utilized to fabricate microtextures on the PDMS elastomer \cite{hill2016surface}. Beyond fluid regulation, the textured surface of PDMS elastomer is also employed in the design of flexible and stretchable sensors for monitoring physiological signals. For example, a study demonstrated the development of a wearable strain sensor, where a laser-texturing technique was used to create microstructures on a composite elastomer made of carbon nanotubes (CNTs) and PDMS \cite{zhou2024laser}. The impact of textured PDMS elastomer extends to cellular behavior, making it highly valuable for regenerative medicine and tissue engineering. One study introduced a nanostructured PDMS elastomer cantilever integrated with silver nanowires (AgNWs-E-PDMS), significantly enhancing cardiomyocyte functionality and offering a promising approach for minimizing the risk of drug-induced cardiotoxicity, particularly during early-stage drug development \cite{liu2023nano}. Furthermore, the textured PDMS elastomer, when integrated with wafer-bonded GaInP/GaAs//Si structures, has been explored to improve the performance of photovoltaic devices \cite{yi2020application}. 

Furthermore, these diverse applications highlight the versatility of textured PDMS elastomers in advancing a wide range of technologies. Consequently, tuning their adhesion behavior is crucial for achieving optimal performance. Despite this importance, there is a notable scarcity of studies specifically addressing interfacial adhesion in systems where the contacting surfaces exhibit random roughness and form complementary interfaces. Most existing research has predominantly focused on periodic or well-defined surface textures. This gap in the literature underscores the need to investigate interfacial adhesion between complementary rough surfaces of PDMS elastomer. In this context, studying the adhesion behavior between complementary randomly rough PDMS surfaces is particularly significant, not only for advancing fundamental understanding of adhesion mechanisms but also for practical applications. This study investigates the effect of surface roughness on the adhesion behavior of PDMS elastomer at a complementary interface. To introduce surface roughness, eight different grit sizes of sandpaper were used, and their topographies were replicated onto the PDMS elastomer through molding. Subsequently, a second replica was fabricated from the initially textured PDMS elastomer surface to form a complementary rough interface. The surface roughness parameters of the PDMS elastomer were quantitatively characterized using a stylus profilometer.
In addition, the interfacial adhesion across these complementary rough PDMS elastomer surfaces was then evaluated using the wedge test method. To further support the experimental findings, the real area of contact was theoretically estimated, providing a correlation with the measured work of adhesion.

\section{Material and methodology}
\subsection{Material}

To fabricate the polydimethylsiloxane (PDMS) elastomer samples, we utilized the commercially available $\mathrm{Sylgard^{TM}}$ 184 kit (Dow Corning), consisting of a PDMS base and a curing agent. The two components were mixed thoroughly in a fixed weight ratio of 10:1 (base to curing agent). $\mathrm{Sylgard^{TM}}$ 184 PDMS elastomer is extensively utilized across various research domains due to its outstanding thermal and chemical stability \cite{ren2020high,brounstein2021long}, mechanical flexibility \cite{an2017pdms,wolf2018pdms}, suitability for micro- and nano-patterning applications \cite{pr8_Micropatterning,pr5_surfacemodification}, biocompatibility \cite{souza2020characterization,pr11_biocompatibility}, excellent corrosion resistance \cite{eduok2017recent}, and high optical transparency \cite{wolf2018pdms}. Owing to its outstanding mechanical properties, chemical properties, and physical properties, $\mathrm{Sylgard^{TM}}$ 184 PDMS elastomer is widely employed in a diverse array of applications across multiple disciplines. Notable examples include its use in wearable sensors \cite{passot2011mechanical}, various coating technologies \cite{gao2020facile, lee2016fabrication, he2018fabrication}, microfluidic devices \cite{nguyen2022multilayer, chen2012photolithographic}, optical components such as lenses \cite{felix2014physical}, soft lithography \cite{chen2012photolithographic, lai2019ratio}, flexible electronics \cite{wiranata2020implementation, li2021flexible}, and soft robotics \cite{wienzek2019elastomeric}.

\subsection{Methodology}
\subsubsection{Preparation of complementary rough surfaces of PDMS elastomer}

\begin{figure*}[!ht]
\centering
  \includegraphics[height=7.7cm]{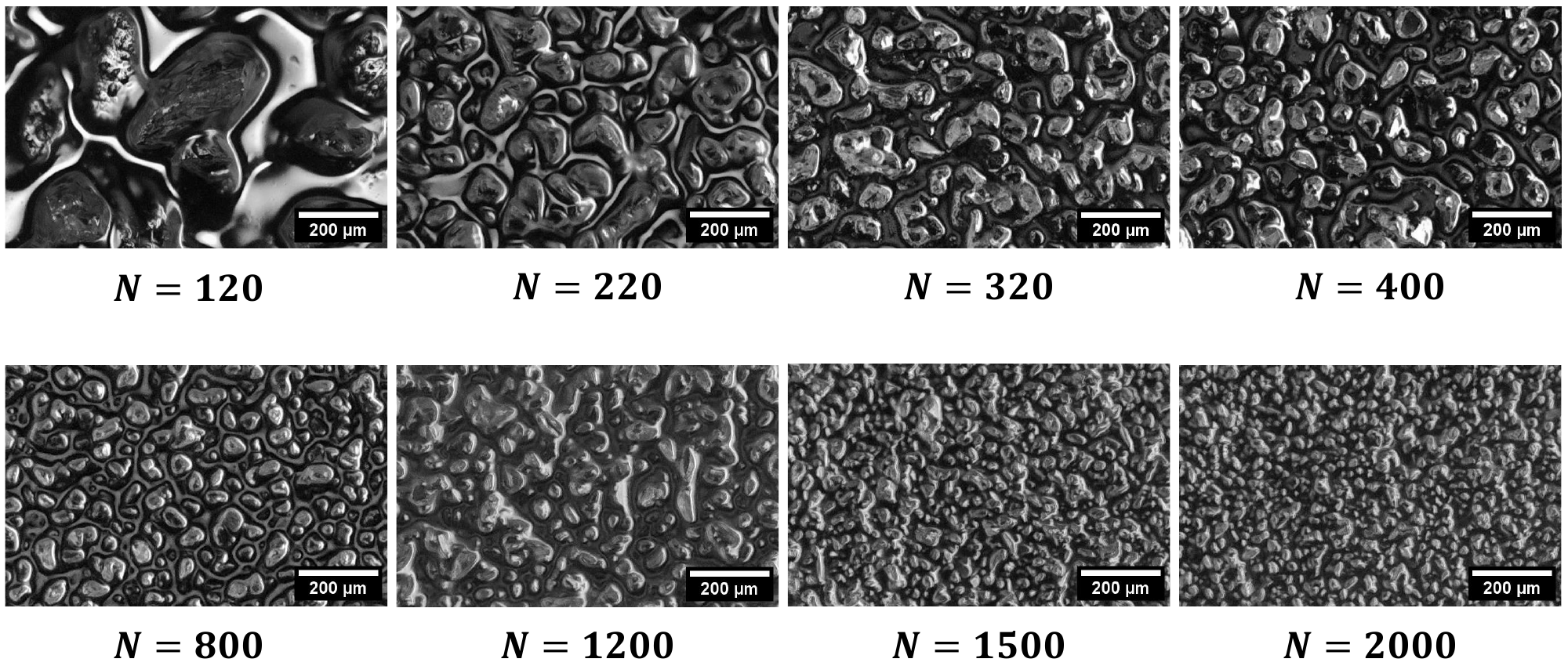}
  \caption{The image shows a rough PDMS elastomer sample fabricated by replicating the surface topography of sandpapers with grit sizes ranging from $N = 120$ to $N = 2000$. The image was captured using an inverted microscope (Olympus IX73, Japan).}
  \label{fgr:sample img}
\end{figure*}

\begin{figure}[!ht]
\centering
\includegraphics[height=5cm]{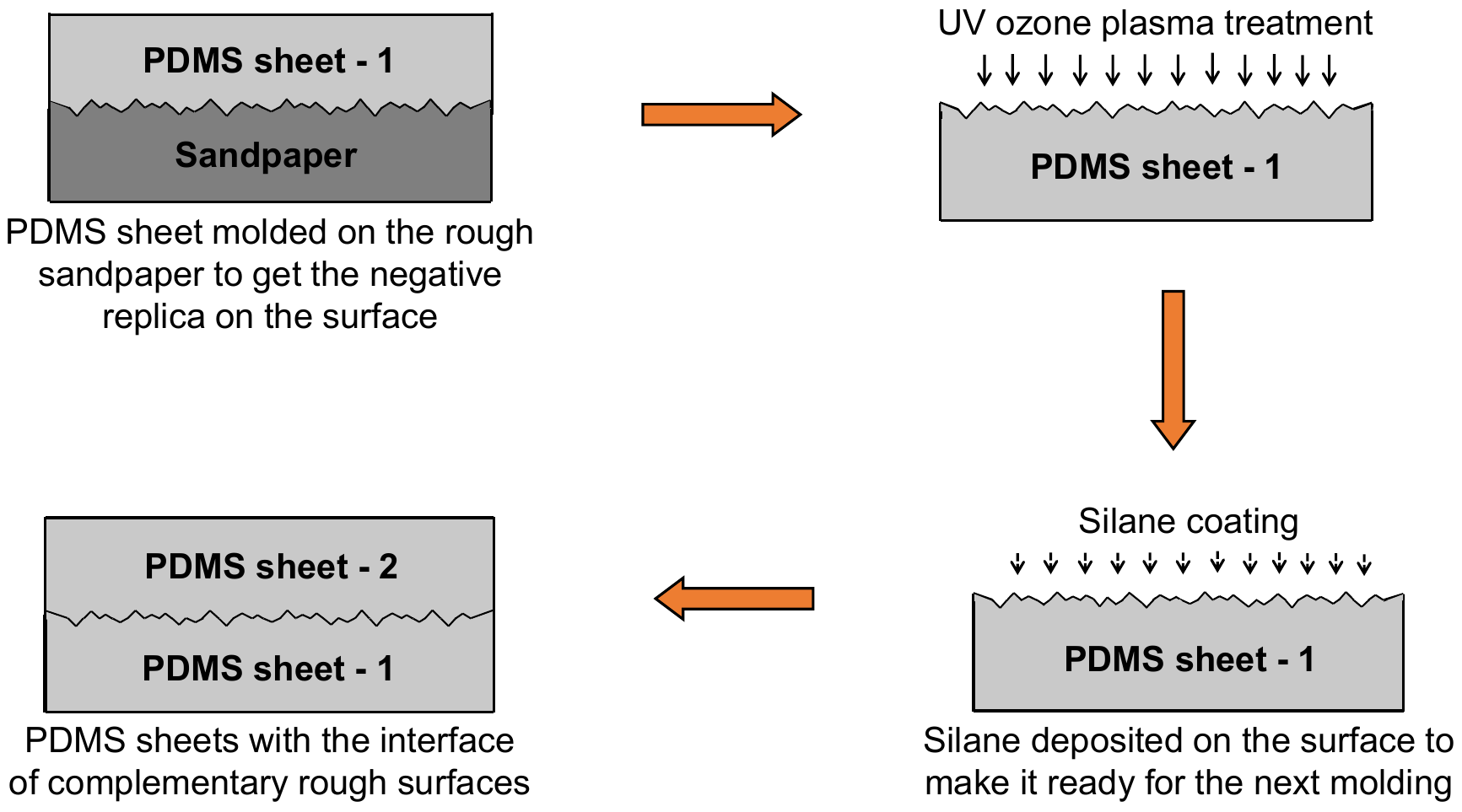}
  \caption{The schematic diagram illustrates the sample preparation process for creating the complementary rough surfaces of the PDMS elastomer.}
  \label{fgr:spp_schm}
\end{figure}

We prepared eight different sets of samples of PDMS elastomer with varying levels of surface roughness as shown in Figure \ref{fgr:sample img}. We used sandpaper (Waterproof, RIKEN CORUNDUM, JAPAN) with various grit sizes ($N$), including 120, 220, 320, 400, 800, 1200, 1500, and 2000, to create surface roughness on the elastomer (PDMS). The complementary rough PDMS elastomer sheets were fabricated following the procedure illustrated in the schematic diagram shown in Figure \ref{fgr:spp_schm}. Initially, the PDMS base was mixed with its curing agent in a beaker at a mixing ratio of 10:1 (w/w) and stirred using a glass rod with a diameter of 6 $\mathrm{mm}$ for approximately 10 to 12 minutes. Following mixing, bubbles formed within the mixture and were subsequently removed using a desiccator under reduced atmospheric pressure, which took approximately 20 to 25 minutes. The degassed mixture was poured onto nitrogen-blown cleaned sandpaper affixed to a glass plate. To ensure uniform thickness of the PDMS elastomer sample, four smooth stainless-steel washers, each approximately 0.88 $\mathrm{mm}$ thick, were placed on the sandpaper. This assembly was covered with acetone (Sisco Research Laboratories Pvt. Ltd.) - cleaned glass plate, and secured using clips. Finally, the samples were cured in an oven at a temperature of  120$^\circ\mathrm{C}$ for a duration of  2 hours. After cooling the sample to room temperature, it was carefully peeled from the sandpaper. The resulting rough PDMS elastomer sheet was designated as PDMS sheet-1. In the subsequent step, a complementary surface of PDMS sheet-1 was fabricated. First, the PDMS sheet-1 was cleaned using a nitrogen blow to remove residual silica particles. The PDMS sheet-1 underwent UV ozone plasma treatment (Advanced Curing System) for 30 minutes. Following this treatment, the ozone-activated surface of PDMS sheet-1 was exposed to n-Octadecyltrichlorosilane $95\%$ (GLR INNOVATIONS) in a desiccator under ambient pressure for 1 hour. Next, a fresh mixture of PDMS base and curing agent was poured onto the silane-treated surface of the PDMS Sheet-1. This assembly was covered with a glass plate and secured with clips. To ensure the desired thickness of the complementary PDMS elastomer sample, four smooth stainless-steel washers ( $\approx$ 1.5 $\mathrm{mm}$ thick) were inserted between the glass plates during fabrication, resulting in a final thickness of $t \approx 0.62$ $\mathrm{mm}$. The sample was then cured in an oven at 120$^\circ\mathrm{C}$ for 2 hours, followed by cooling to room temperature. The complementary rough PDMS elastomer sheet (PDMS Sheet-2) was then carefully peeled from PDMS Sheet-1, with only 80$\%$ of the total area being separated.  Finally, the PDMS sheet-1 and PDMS sheet-2 were subjected to thermal treatment in the oven at a temperature of 150$^\circ\mathrm{C}$ for 10 minutes to remove any residual silane layer. Subsequently, the sheets were cleaned using a nitrogen blow. The samples were then precisely cut into the dimensions of $4 \mathrm{cm}\times1\mathrm{cm}$. The final measured thicknesses of PDMS sheet-1 and PDMS sheet-2 were approximately 0.88 $\mathrm{mm}$ and 0.62 $\mathrm{mm}$, respectively.

\subsubsection{Surface roughness characterisation}
Surface roughness measurements were performed using the contact method with a stylus profilometer (Dektak XT, BRUKER). For transparent materials and surfaces with varying scales of isotropic roughness, the contact method generally offers advantages over non-contact techniques, such as optical methods \cite{rodriguez2025determine}. When the scan length is adequate, this line-based measurement approach provides comprehensive and reliable data on a wide range of surface roughness characteristics. In this method, the height and depth of surface roughness are assessed through mechanical contact, wherein a stylus moves across the peaks and valleys of the sample surface under a minimal contact load. The position of the stylus is primarily determined by its repulsive interaction with the sample surface. This method exhibits high sensitivity to surface roughness with minimal interference. The vertical displacement of the stylus is converted into an electrical signal by a transducer, which subsequently provides the profile height $z(x)$ of the surface roughness. The stylus tip radius, scan length, and stylus force were set to 2 ${\mu{m}}$, 1500 $\mu{m}$, and 2 $\mathrm{mg}$, respectively. The stylus profilometer has a vertical resolution of 0.1 $\mathrm{nm}$, while its horizontal resolution is limited by the diameter of the stylus tip. Multiple scans were conducted at various positions on the sample, and the root mean square (RMS) roughness ($R_q$) was determined by selecting data from three of these scans. The root mean square (RMS) roughness ($R_q$), is defined by the following expression \cite{gadelmawla2002roughness}:

\begin{equation} \label{eq_(1)}
R_q = \sqrt{{\frac{1}{L}\space\int_{0}^{L} z(x)^2 \, dx}}
\end{equation}
where $L$ is the scan length, and $z(x)$ shows the profile height of the roughness at a scan distance $x$ from the mean line.

\subsubsection{Evaluation of real contact area}

To calculate the real area of contact $A_r$, we have considered three different cases. In the first case, the asperities are assumed to have a hemispheroidal shape. In the second case, the asperities are modeled as conical. In the third case, the asperities are considered to be a combination of both hemispheroidal and conical shapes, with a diameter of $d$ and a root mean square roughness of $R_q$.

In the first case, i.e., for hemispheroidal asperities, the real area of contact for $n$ number of asperities on the surface can be expressed as follows:

\begin{equation} \label{eq_(2)}
A_{r(s)} = {n} {\pi} {d}\space[\frac{(d/2)^{1.6} + 2\times R^{1.6}_q}{3}]^{\frac{1}{1.6}}
\end{equation}

In the second case, i.e., for conical asperities, the real area of contact for $n$ number of asperities on the surface can be expressed as follows:

\begin{equation} \label{eq_(3)}
A_{r(c)} = \frac{{n}{\pi}{d}}{2}\space[(d/2)^{2} + R^{2}_q]^{\frac{1}{2}}
\end{equation}

In the third case, i.e., for a combination of both hemispheroidal and conical-shaped asperities, the real area of contact for $n$ number of asperities on the surface can be expressed as follows:

\begin{equation} \label{eq_(4)}
A_{r(s+c)} = {\frac{1}{2}}[A_{r(s)} + A_{r(c)}]
\end{equation}

To evaluate the diameter of asperities on various randomly rough surfaces, high-resolution images of the rough PDMS elastomer sample were captured using a microscope (IX 73, Olympus, Japan). Additionally, open-source software (ImageJ) was utilized to extract the asperity diameters. A total of 100 data points were collected from various locations on the sample surface, and the average of these data points was calculated to determine the asperity diameter.

\subsubsection{Wedge test approach for adhesion measurement}

Elastomer (PDMS) sheets with a randomly rough interface were manually peeled apart, resulting in two sheets, each possessing complementary surface roughness. Adhesion at the interface of these complementary rough surfaces was evaluated as follows. First, the two complementary peeled PDMS elastomer sheets were manually pressed together to ensure attachment. To achieve near-perfect alignment of the sheets, only 80$\%$ of the total area was peeled. It is important to note that applying sufficient pressure was necessary to ensure proper attachment of the complementary PDMS elastomer sheets. Elastomer (PDMS) sheets with high root mean square roughness $(R_q)$ values also exhibit larger asperity diameters $(d)$, leading to shallower valleys. Consequently, these complementary PDMS elastomer sheets require significantly less pressure for effective conformal attachment. After the removal of the applied pressure from the complementary attached PDMS elastomer sheets, spontaneous detachment was observed at certain points of the interface. This detachment may have occurred due to local mismatches between the peaks and valleys of the complementary PDMS elastomer sheets. However, this local mismatch between peaks and valleys never leads to the complete detachment of the complementary PDMS elastomer sheets. Additionally, no time-dependent effects were detected in the detachment of the PDMS elastomer sheets following the removal of pressure.

\begin{figure}[!ht]
\centering
  \includegraphics[height=5.9cm]{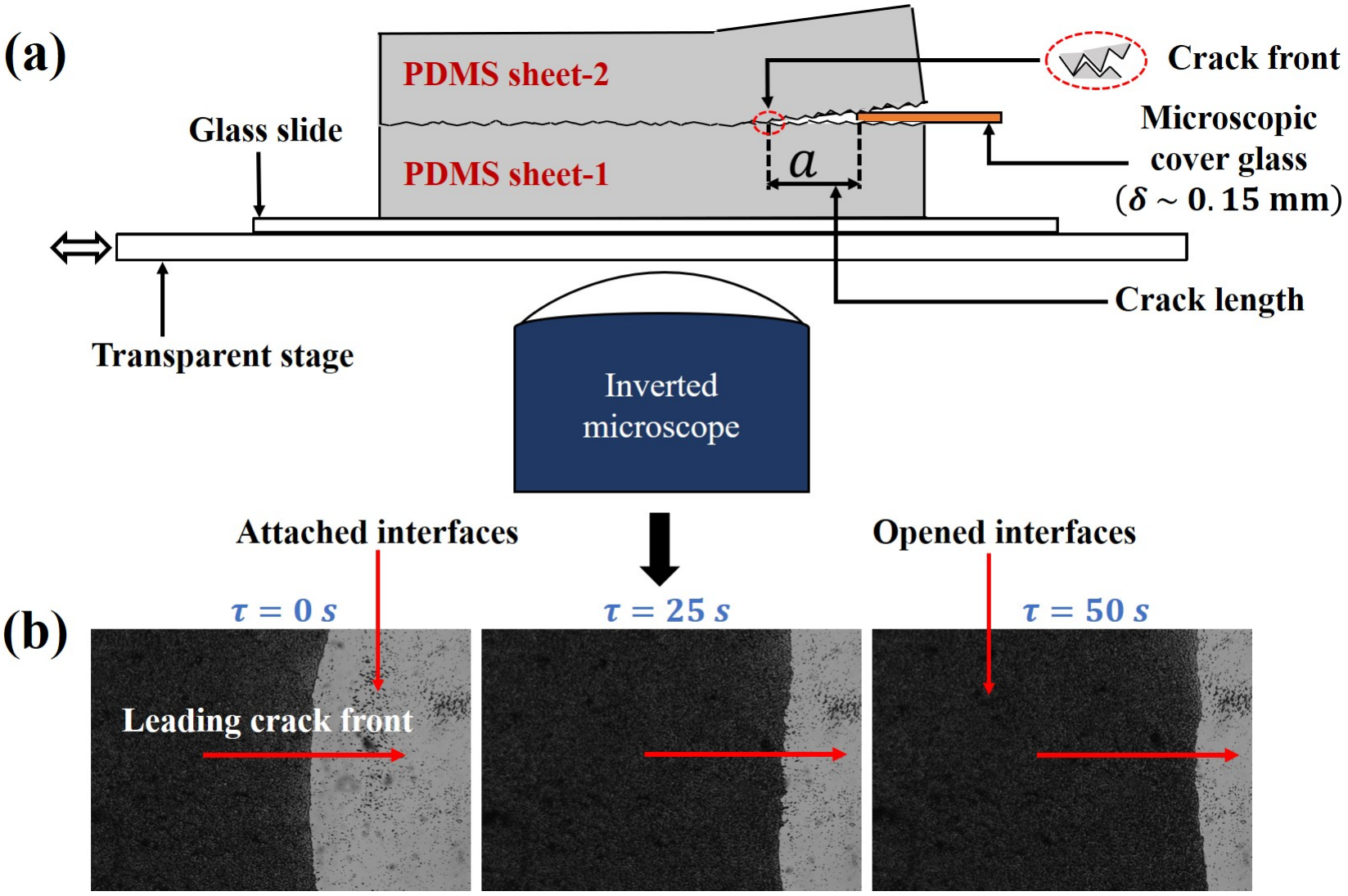}
  \caption{(a) The diagram showing the schematic of the experimental setup integrated into the inverted optical microscope (IX 73, Olympus, Japan). (b) The image shows the crack propagation recorded at distinct time intervals. The arrowhead indicates the surface in contact, while the tail points to the open surface.}
  \label{fgr:exp_setp}
\end{figure}

Figure \ref{fgr:exp_setp} (a) presents a detailed schematic diagram of the wedge test experiment. The entire setup for the wedge test was assembled using an inverted microscope (Olympus IX 73, Japan). Initially, the complementary PDMS elastomer sheets were placed on a glass slide and subsequently fixed onto the transparent stage of the inverted microscope using crystal-clear tape (kangaro). Furthermore, a microscopic cover glass (BLUE STAR) with a thickness of $\delta \approx 0.15 \space \mathrm{mm}$ was inserted at the interface of the rough complementary PDMS elastomer sheets to form a wedge, thereby initiating a crack at the interface. The crack front and the corresponding crack length $(a)$ were utilized to illustrate the crack growth at the interface. Subsequently, a vertical force was applied to the top of the two halves of the complementary PDMS elastomer sheets to minimize the crack length to its lowest value. Upon the removal of the applied force, crack propagation commenced immediately along the length of the sample at the interface, driven by the spontaneous initiation of the crack front at the edge of the microscopic cover glass. The propagation of the crack was monitored using a high-speed camera (Nikon) integrated with an inverted microscope. The camera operated at a capture rate of 100 frames per second. A commercial software, View7, provided with a camera, was used to record the crack propagation videos. The recorded videos were subsequently analyzed using Tracker, an open-source video-based motion analysis software. A tracker was utilized to determine the crack propagation direction, crack length, and crack propagation velocity \cite{c22_brown2009innovative,c23_moraru2021distance,c24_rodrigues2013teaching,c25_wee2012using}. From the recorded videos, we extracted the raw data of incremental crack length over time. These raw data were then used to determine the crack propagation velocity $(v)$ and energy release rate $(G_r)$. Subsequently, we evaluated the equilibrium crack length $(a_e)$ (the point at which crack propagation velocity reaches zero) \cite{kumar2025adhesion,kumar2025controlling,r41_Chiranjit} and the threshold value of the work of adhesion for different sets of rough complementary PDMS elastomer samples.

Fig. \ref{fgr:exp_setp} (b) illustrates the propagation of the crack front using three distinct optical images extracted from the Tracker software at different time intervals: $\tau=0\space{s}$, $\tau=25\space{s}$, and $\tau=50\space{s}$. The region behind the leading crack front represents the open interface, indicating that a crack has already formed in this area. Conversely, the region ahead of the leading crack front corresponds to the attached interface, signifying that the crack has not yet developed in this section.\par

To investigate the effect of surface roughness on the adhesion behavior of PDMS elastomers, a simplified theoretical model was employed. In this model, two PDMS elastomer strips, characterized by identical surface roughness profiles and exhibiting complementary geometries, are brought into contact. The assumption of conformal contact between the surfaces allows us to neglect the presence of any residual stresses that might arise from interfacial mismatches or imperfections. The composite system formed through the conformal contact of PDMS elastomer sheets can be idealized as an infinite elastic medium with elastic homogeneity in the bulk. However, the contact surface between the sheets constitutes a weak internal interface, arising from surface roughness and reduced mechanical strength, thereby representing a localized zone of weakness within the material.
In this analysis, we assume that the roughness parameters of the crack surface are small in comparison to characteristic dimensions like the overall crack length and the specimen size. This simplification enables a focused investigation of the local behavior near the semi-infinite crack tip, which is subjected to remote Mode I (opening mode) loading.
Under these conditions, and in the absence of undulating geometries or features that would guide crack propagation along the weak interface, the crack is expected to propagate perpendicular to the direction of the applied tensile load. This behavior arises from the fact that the energy release rate, a fundamental parameter governing crack growth, reaches its maximum in this orientation. Consequently, the crack advances along the path that maximizes energy dissipation, as predicted by the principle of maximum energy release. This direction of propagation aligns with that observed in media characterized by flat, homogeneous interfaces, thereby indicating that, when geometric undulations and material inhomogeneities are absent, the crack path is primarily governed by far-field loading conditions and the intrinsic elastic response of the medium.
  
In the context of Mode I fracture, the energy release rate $G_r$ under conditions of fixed displacement can be formally expressed as follows:

\begin{equation} \label{eq_(5)}
G_r = \frac{3E^*\delta^2t^3}{4a^4}
\end{equation}

Given that $u=2\delta$, where $\delta$ denotes the thickness of the cover glass, $a$ represents the crack length, and $t$ is the thickness of the PDMS elastomer sheet. The plane strain modulus, denoted as $E^*$, is given by
$E^* = \frac{E_{\mathrm{PDMS}}}{(1-\nu_{\mathrm{PDMS}}^2)}$, where $E_{\mathrm{PDMS}}$ and $\nu_{\mathrm{PDMS}}$ are modulus of elasticity and Poisson’s ratio of the PDMS elastomer, respectively.\par

In the case of quasi-static crack propagation along a flat interface or during unguided interfacial crack propagation, the work of adhesion $(W_{\mathrm{ad}})$ is coupled with the energy release rate. For further details, refer to \cite{kumar2025adhesion, kumar2025controlling}. The following expression represents the correlation between the work of adhesion and crack length. 
 
\begin{equation} \label{eq_(6)}
W_{\mathrm{ad}} = \frac{3E^*\delta^2t^3}{4a^4}
\end{equation}

At the point of equilibrium, where crack propagation ceases, and the crack velocity $(v)$ becomes zero, the corresponding equilibrium crack length is given by $a = a_e$, and the threshold value of the work of adhesion is denoted as $W_{\mathrm{ad}}^0$. The following expression provides the relationship between these two quantities:

 \begin{equation} \label{eq_(7)}
W^0_{\mathrm{ad}} = \frac{3E^*\delta^2t^3}{4a_e^4}
\end{equation}

The equation \eqref{eq_(7)} thus illustrates the relationship between the equilibrium crack length and the threshold value of the work of adhesion at the interface of the complementary surface of the PDMS elastomer.

\section{Results and discussion}

\begin{figure}[!ht]
    \centering
    \includegraphics[height = 6.8cm]{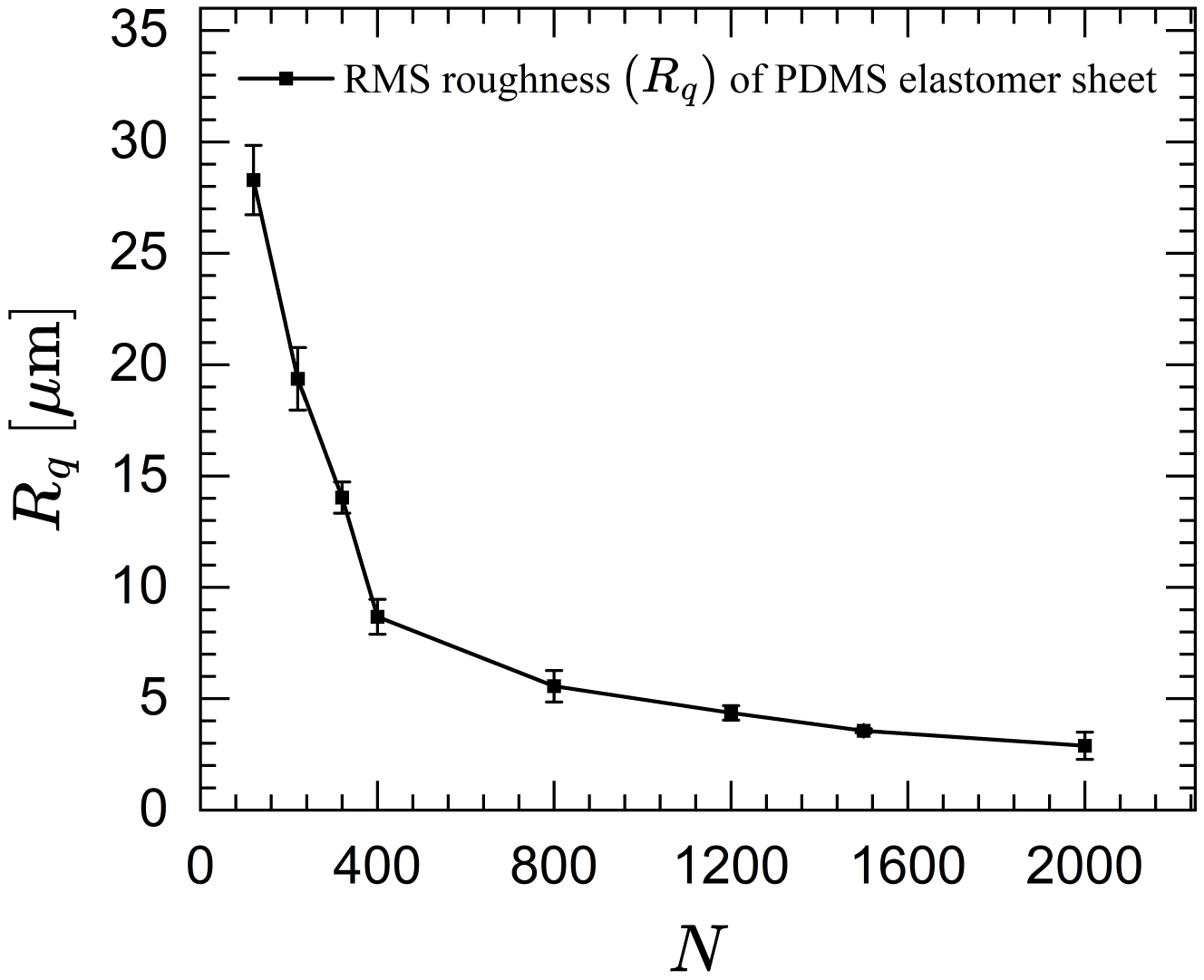}
    \caption{Illustrate the relationship between the root mean square roughness ($R_q$) of the PDMS elastomer sample and the grit sizes of different sandpapers ($N$).}
    \label{fig: ink_Rq_N}
\end{figure}

Figure \ref{fig: ink_Rq_N} illustrates the correlation between the root mean square roughness ($R_q$) of PDMS elastomer surfaces and the grit sizes ($N$) of different sandpapers. The grit sizes range from $N=120$ to $N = 2000$. The data provide insights into how surface roughness varies with abrasive particle size, highlighting the influence of grit size on the topographical characteristics of the PDMS elastomer surface.
We utilized the contact method with a stylus profilometer to analyze the surface roughness of the PDMS elastomer. This technique involves a fine stylus traversing the surface of the material, capturing variations in height to quantify roughness parameters. This technique enables precise surface characterization by detecting fine variations in roughness. Additionally, its ability to measure both micro- and nanoscale surface features ensures accurate quantification of surface irregularities.
The stylus profilometer provides high-resolution measurements and can be effectively used for transparent materials, making it a reliable method for assessing the topographical features of PDMS elastomer. For a given grit size of sandpaper, variations in the roughness pattern can occur due to the inherent randomness in the microstructure of the abrasive particles. Consequently, even sandpapers with the same nominal grit size may exhibit differences in surface topography, affecting the roughness of the material. The surface roughness of the PDMS elastomer sample was measured at multiple locations. To ensure accurate and representative reporting, three data points exhibiting the most consistent and characteristic roughness values were selected for the calculation and presentation of the root mean square roughness $R_q$. We observed that the roughness $R_q$ of the PDMS elastomer exhibited an inverse relationship with the grit size of the sandpaper. Specifically, as the grit size increased, the measured surface roughness decreased. This trend can be attributed to the fact that sandpapers with higher grit sizes contain finer abrasive particles, resulting in smoother surface textures on the PDMS elastomer. Conversely, lower grit sizes correspond to coarser abrasives, leading to increased surface roughness.

\begin{figure}[!ht]
    \centering
    \includegraphics[height = 7cm]{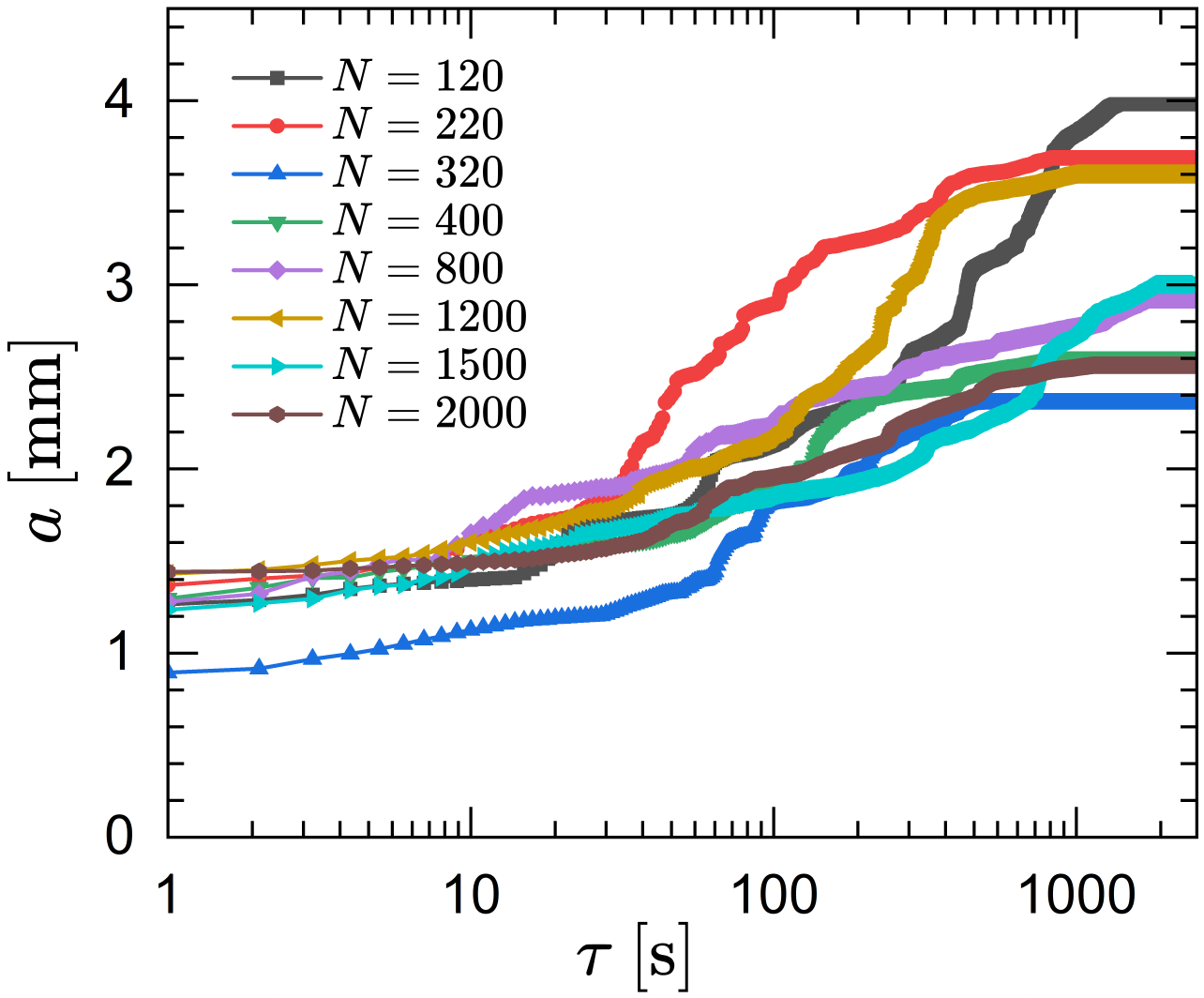}
    \caption{The graph illustrates the variation of crack length $(a)$ as a function of time $(\tau)$. Each curve in the graph corresponds to a different grit size  $(N)$, ranging from $N=120$ to $N=2000$.}
    \label{fig:ink_at}
\end{figure}

Figure \ref{fig:ink_at} illustrates the relation between crack length ($a$) and time ($\tau$) for different grit sizes, including $N=120$, $N=220$, $N=320$, $N=400$, $N=800$, $N=1200$, $N=1500$, and $N=2000$. Each curve in the figure represents a specific grit size, facilitating a comparative analysis of the data. The observed trends offer valuable insights into the influence of surface roughness on the progression of crack growth. The initial vertical displacement ($u = 2\delta$) is consistent across all samples; however, the initial crack length varies. This variation arises due to differences in the starting point of data acquisition for each sample. Specifically, at the onset of the experiment, crack propagation occurs at a rapid rate, making it challenging to capture a uniform initial measurement across all samples. Over time, the propagation rate decreases, eventually reaching a stage where data tracking becomes more consistent. However, the time at which this transition occurs varies between samples, leading to differences in the recorded initial crack length. The crack length increases over time; however, the variation in crack length observed in each curve is non-uniform. This non-uniformity arises due to the roughness present at the interface, which alters the direction of crack propagation. As the crack encounters surface irregularities, its trajectory deviates. These directional changes affect the magnitude of the crack propagation rate in different segments of the curve, resulting in fluctuations in the overall growth behavior. The crack growth ceases after a certain period, referred to here as the equilibrium point. At this stage, the contacting surfaces are completely separated, and the crack reaches a stable length, beyond which no further propagation occurs. The crack length associated with this equilibrium state represents a threshold value termed the equilibrium crack length ($a_e$). 
For each sample, the equilibrium crack length exhibits a non-monotonic variation with respect to grit size ($N$). Specifically, within the grit size range of $N=120$ to $N=320$, The equilibrium crack length decreases. However, as the grit size increases from $N=320$ to $N=1200$, the equilibrium crack length shows a rising trend. Beyond $N=1200$, the equilibrium crack length decreases once again. Notably, within the range of $N=120$ to $N=2000$, the minimum equilibrium crack length is observed at $N=320$, while the maximum equilibrium crack length is recorded at $N=1200$. The crack propagation velocity corresponding to this threshold value of crack length is zero since there is no crack propagation occurring beyond this point. The critical or threshold value of the work of adhesion corresponding to this zero crack propagation velocity is simply referred to here as the work of adhesion ($W^0_\mathrm{ad}$). The work of adhesion is directly influenced by the equilibrium crack length. Specifically, it increases as the equilibrium crack length decreases and decreases as the equilibrium crack length increases. This relationship is analyzed and discussed in detail in the following sections.

\begin{figure}[!ht]
    \centering
    \includegraphics[height = 6.9cm]{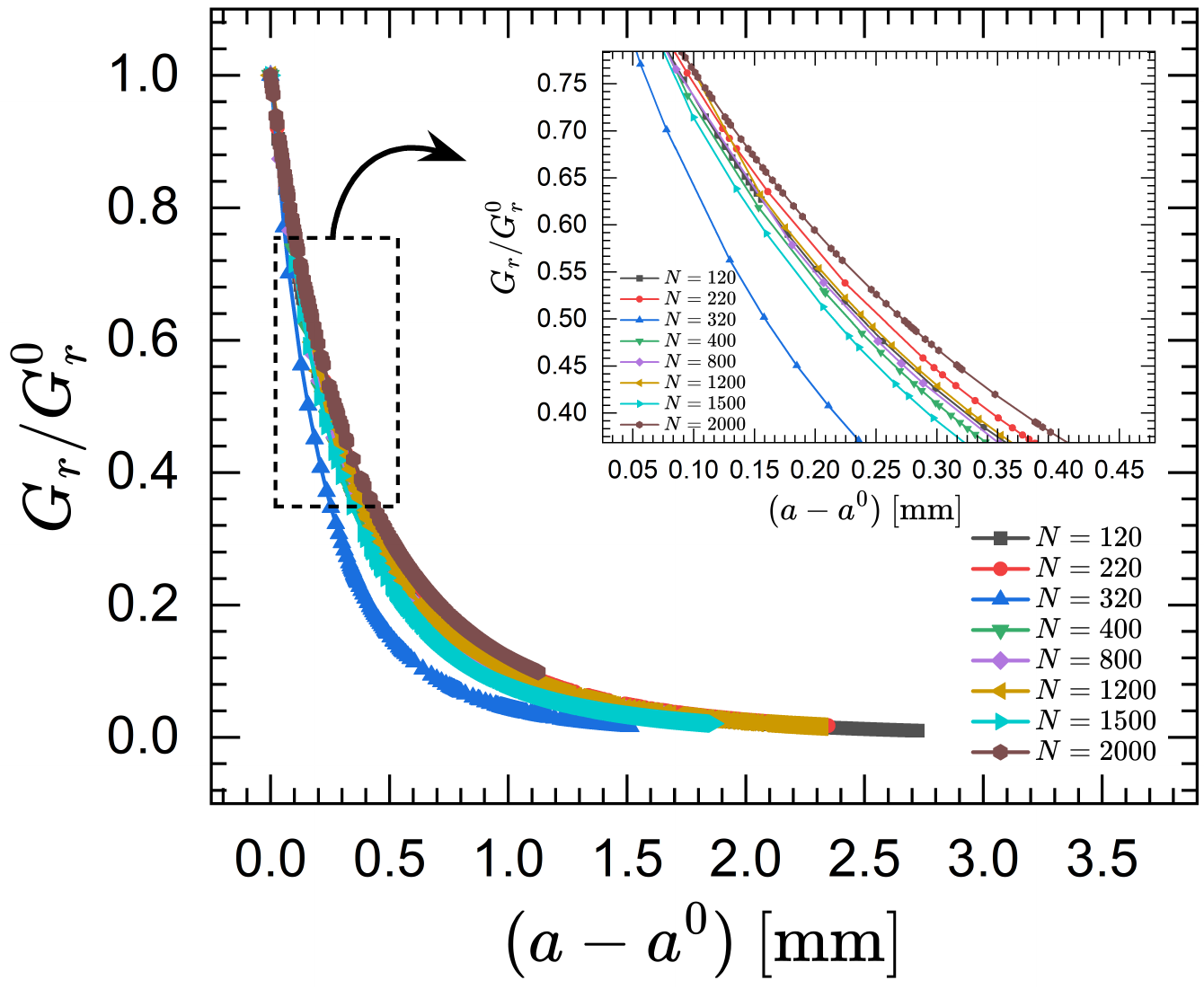}
    \caption{The graph illustrates the relationship between the normalized energy release rate ($G_r/G^0_r$) and crack length ($a-a^0$). Each curve in the graph corresponds to a different grit size ($N$). The inset graph within the plot highlights the decreasing trend of the normalized energy release rate, providing a more detailed view of this trend.}
    \label{fig:ink_Ga}
\end{figure}

Figure \ref{fig:ink_Ga} presents the relationship between the normalized energy release rate, expressed as ($G_r/G^0_r$), and the crack length increment, ($a-a^0$), for the samples of different grit sizes ($N$). The normalized energy release rate is defined as the ratio of the energy release rate ($G_r$) at a given instance to its initial value ($G^0_r$), which corresponds to the energy release rate at the time $\tau=0\space\mathrm{s}$. This normalization enables a comparative analysis of the energy release rate behavior during crack propagation across samples with varying grit sizes. The initial crack length, denoted as $a^0$, is the crack length measured at the time $\tau=0 \space\mathrm{s}$. This energy release rate directs the progression of crack growth and influences both the rate and direction of crack propagation at the interface of the contacting surfaces. At the start of crack propagation at the interface of the complementary surfaces of the PDMS elastomer, the $G_r/G^0_r$ is initially high. As the crack advances along the length of the sample, the $G_r/G^0_r$ decreases rapidly. However, as the crack approaches the equilibrium point, the rate of decrease slows down. Upon reaching equilibrium, the $G_r/G^0_r$ stabilizes and attains a constant value. This behavior suggests that crack propagation undergoes a transition from an unstable to a stable regime, where the driving energy for further crack growth diminishes. This trend in the variation of the $G_r/G^0_r$ is observed consistently across all samples, regardless of the grit size. However, the $G_r/G^0_r$ at the equilibrium point is not consistent across all samples due to the inconsistent variation in the threshold value of ($a-a^0$). This inconsistency arises because the initial crack length $a^0$ varies among different samples.

\begin{figure}[!ht]
    \centering
    \includegraphics[height = 6.9cm]{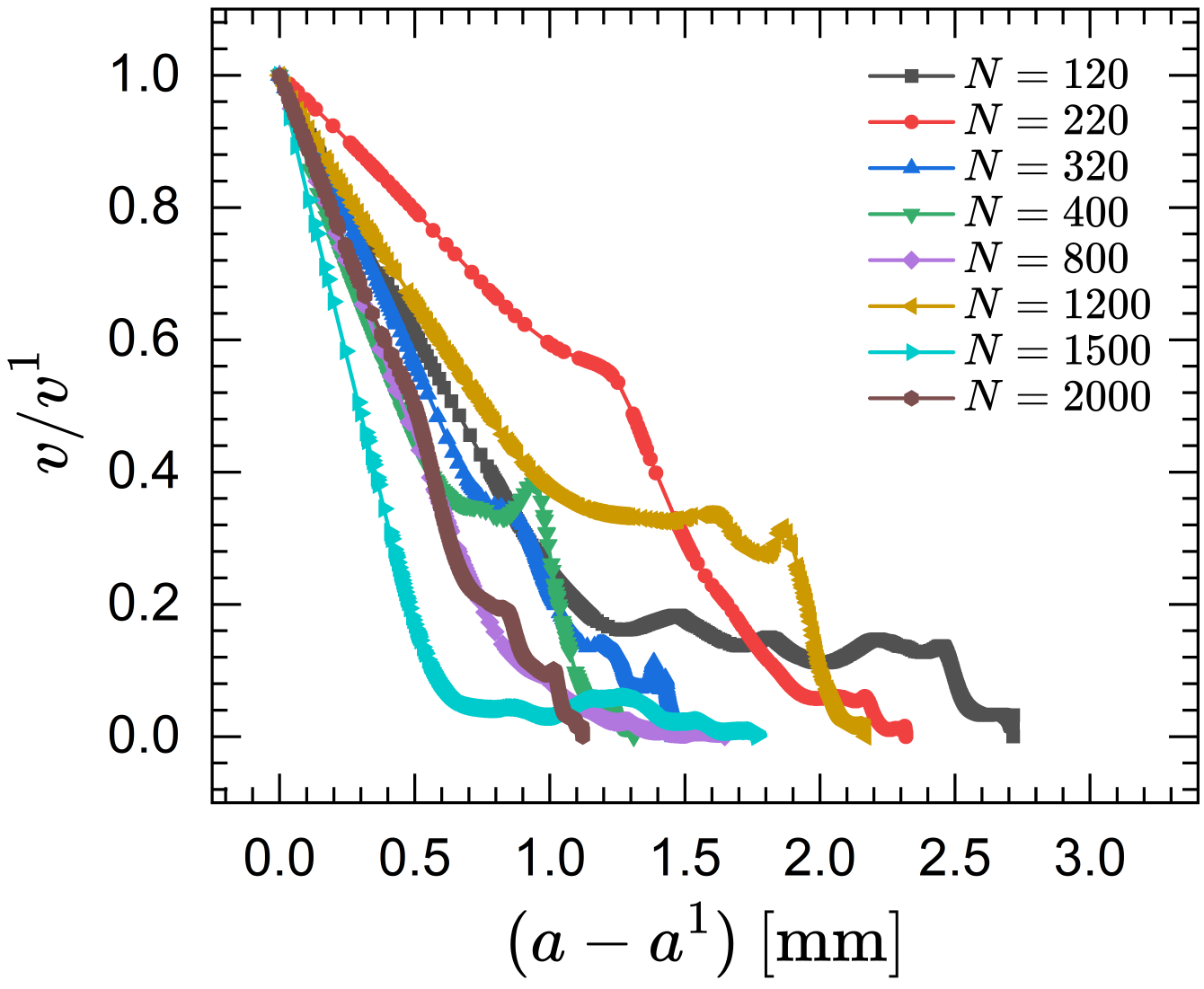}
    \caption{The graph shows the relationship between the normalized crack propagation velocity ($\upsilon/\upsilon^1$) and the crack length ($a-a^1$) corresponding to the samples of different grit sizes ($N$).}
    \label{fig:ink_va}
\end{figure}

Figure \ref{fig:ink_va} represents the correlation between the normalized crack propagation velocity ($\upsilon/\upsilon^1$) and crack length ($a-a^1$) for samples with various grit sizes ($N$). The ratio ($\upsilon/\upsilon^1$) is defined as the crack propagation velocity ($\upsilon$) at a given instant $\tau$ divided by the reference velocity ($\upsilon^1$), which is the crack propagation velocity at $\tau=1\space\mathrm{s}$. This ratio enables a comparative analysis of velocity variations across different samples of PDMS elastomer subjected to various grit sizes. The parameter $a^1$ represents the crack length at the reference time $\tau=1\space\mathrm{s}$. To analyze the progression of the crack at the interface of the complementary surfaces of the elastomer (PDMS), we have considered $(\tau = 0 \mathrm{s})$ as the initial reference point. At this moment, a finite initial crack length, denoted as $(a^0)$, was observed. However, the crack propagation velocity remained zero. As time progressed to $(\tau = 1 \mathrm{s})$, both the crack propagation velocity and the crack length attained a constant value, depicted as $(\upsilon^1)$ and $(a^1)$, respectively. Given that the crack propagation velocity was initially zero at $(\tau = 0 \mathrm{s})$, we chose to normalize the velocity with respect to $(\upsilon^1)$ rather than $(\upsilon^0)$ ensuring a meaningful and consistent basis for comparison of velocity variations over time. The ratio ($\upsilon/\upsilon^1$), which indicates the normalized crack propagation velocity, decreases as the crack length ($a-a^1$) increases. However, to reduce fluctuations in $\upsilon/\upsilon^1$ and enhance the interpretability, a smoothing average (Lowess) has been applied to each curve.
At the onset of crack propagation along the interface of the complementary surface of the PDMS elastomer, the decrease in the normalized crack propagation velocity, $\upsilon/\upsilon^1$, is initially rapid. This is due to the high driving energy available for crack propagation at the early stages. As the crack length reaches equilibrium, the rate of decrease in $\upsilon/\upsilon^1$ gradually slows down. Once the equilibrium crack length is reached, the normalized crack propagation velocity ultimately reduces to zero, as the driving energy is fully dissipated and no longer available to sustain further crack growth. The trend of variation in $\upsilon/\upsilon^1$ with respect to ($a-a^1$) along the sample interface remains consistent across all samples, despite the different grit sizes. Furthermore, the ratio $\upsilon/\upsilon^1$ at the equilibrium point is not uniform across all samples. This inconsistency arises because the crack length $a^1$, corresponding to the time $(\tau = 1 \mathrm{s})$, varies among different samples. As a result, the threshold value of ($a-a^1$) exhibits non-uniform variation, which in turn leads to inconsistencies in the ratio $\upsilon/\upsilon^1$ at the equilibrium point across all the samples.

\begin{figure}[!ht]
    \centering
    \includegraphics[height = 6.7cm]{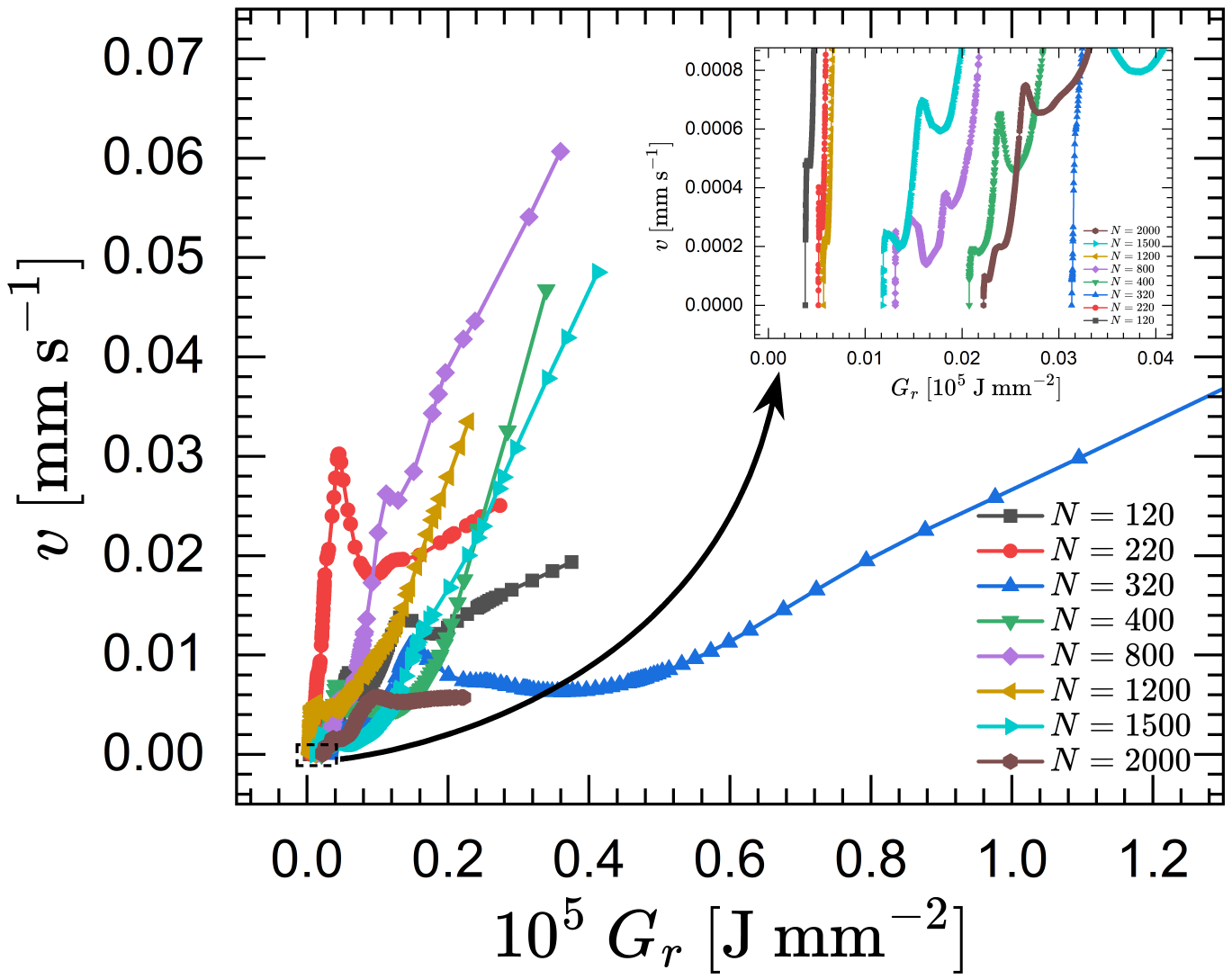}
    \caption{The plot shows the relationship between crack propagation velocity ($\upsilon$) and the energy release rate ($G_r$), with each data set corresponding to a distinct grit size ($N$) of the sandpaper.}
    \label{fig:ink_Gv}
\end{figure}

Figure \ref{fig:ink_Gv} 
illustrates the relationship between crack propagation velocity ($\upsilon$) and the energy release rate ($G_r$), with each data set corresponding to different sandpaper grit sizes ($N$). As previously discussed in Fig. \ref{fig:ink_at}, at the equilibrium point, crack propagation ceases, and the crack propagation velocity reduces to zero. This condition defines the threshold crack length, also referred to as the equilibrium crack length ($a_e$), as well as the threshold value of the work of adhesion ($W^0_\mathrm{ad}$). The vertical displacement of the crack remains fixed across all samples. As a result, the initial energy release rate and crack propagation velocity are high in all cases. However, as the cracks begin to propagate at the interface in a direction perpendicular to the crack front, both the energy release rate and crack propagation velocity gradually decrease. This reduction continues until the system reaches an equilibrium state, at which point the energy release rate stabilizes at a constant value, and the crack propagation velocity ultimately reduces to zero. In certain curves, the fluctuation in crack propagation velocity appears to be significant. This variation arises from the presence of unattached contact points at certain locations along the sample interface. At these specific points, the crack length undergoes a sudden increase, leading to a sharp rise in crack propagation velocity. Consequently, the velocity deviates from a consistent trend and instead exhibits irregular fluctuations. However, to mitigate these effects and enhance interpretability, a smoothing average (Lowess) has been applied to each curve. When comparing the threshold values of the energy release rate corresponding to zero crack propagation velocity, a non-monotonic trend is observed, as shown in the inset figure, with variation in grit size within the range of $N=120$ to $N=2000$. Specifically, the threshold energy release rate increases from $N=120$ to $N=320$, reaching its first peak at $N=320$. Beyond this point, it decreases as $N$ increases from $N=320$ to $N=1200$, where it attains its minimum value within the examined range. Subsequently, the threshold energy release rate begins to increase again from $N=1200$ to $N=2000$. This pattern suggests the existence of two optimum points within the studied range of grit sizes. The first optimum occurs at $N=320$, where the threshold energy release rate reaches its maximum value, while the second occurs at $N=1200$, where the threshold energy release rate attains its minimum.

 \begin{figure}[!ht]
    \centering
    \includegraphics[height = 6cm]{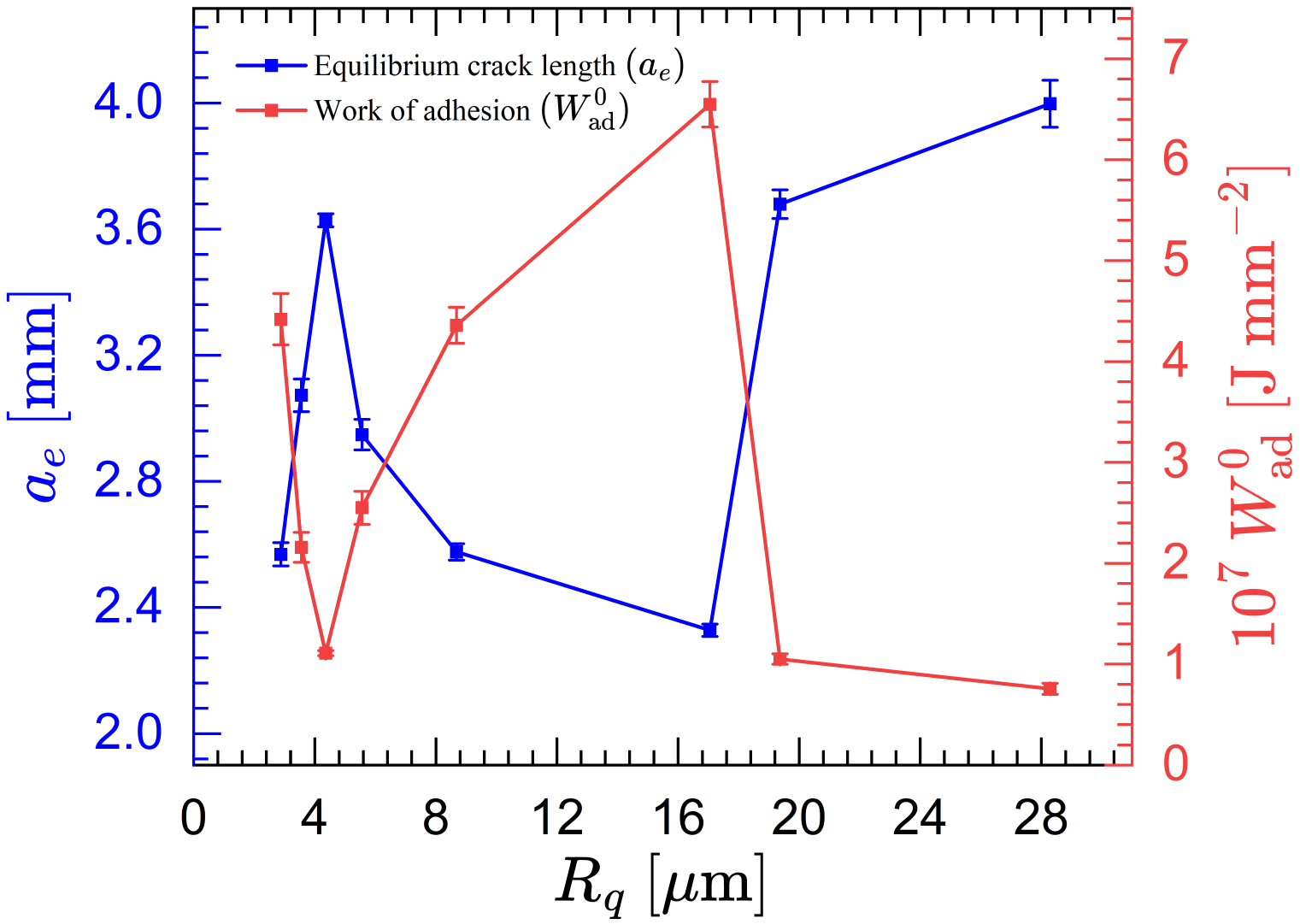}
    \caption{The graph illustrates the relationship between the equilibrium crack length ($a_e$) and the work of adhesion ($W^0_\mathrm{ad}$) as functions of root mean square (RMS) roughness ($R_q$) for samples with different grit sizes ($N$) ranging from $N=120$ to $N=2000$.}
    \label{fig:ink_a0_Wad}
\end{figure}

\begin{figure*}[!ht]
    \centering
    \includegraphics[height = 6cm]{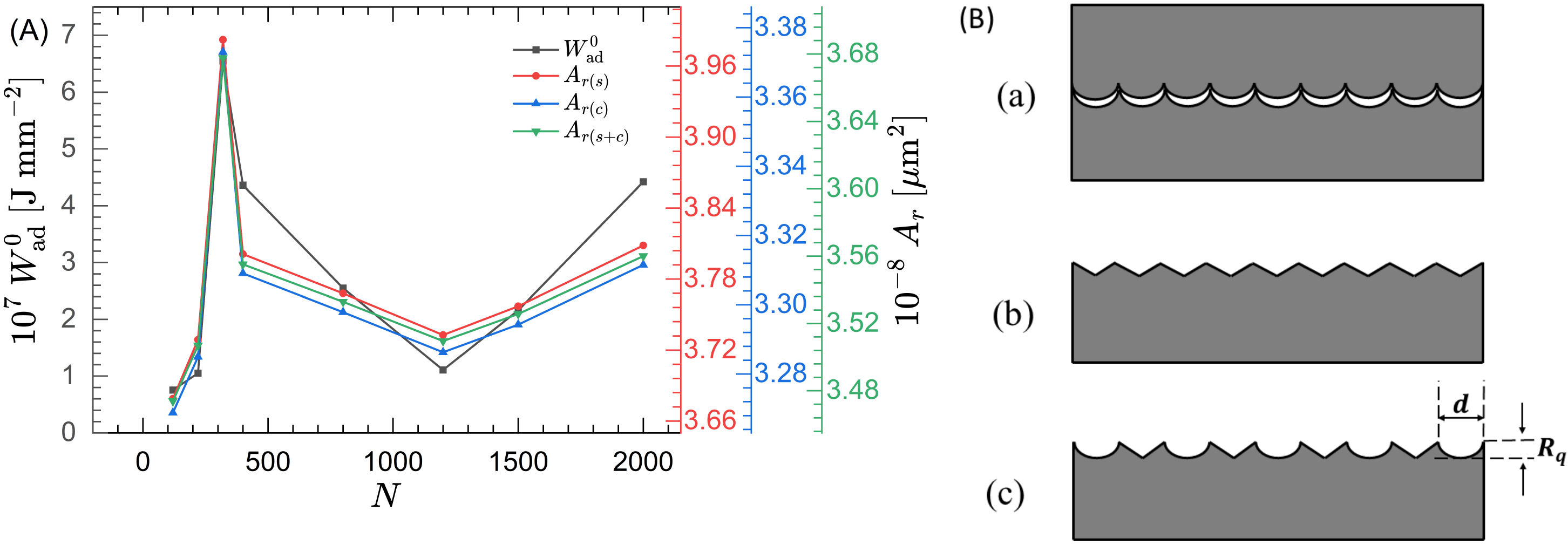}
    \caption{(A) The graph illustrates the variation in the real area of contact ($A_r$) and work of adhesion ($W^0_\mathrm{ad}$) as a function of grit size ($N$). The symbols $A_{r(s)}$, $A_{r(c)}$, and $A_{r(s+c)}$ represent the real area of contact corresponding to hemispheroidal asperities, conical asperities, and a combination of both hemispheroidal and conical-shaped asperities, respectively. (B) Illustrates the schematics of three different types of asperities. (a) A hemispheroidal shape of asperities at the interface. (b) A conical shape of asperities at the interface. (c) A combination of both hemispheroidal and conical shapes of asperities at the interface.} 
    \label{fig:ink_Wad_Ar_N}
\end{figure*}

Figure \ref{fig:ink_a0_Wad} presents the final output graph depicting the equilibrium crack length ($a_e$) and work of adhesion ($W^0_\mathrm{ad}$) as functions of root mean square (RMS) roughness ($R_q$) for samples with different grit sizes ($N$). Earlier in the graph of crack length ($a$) versus time ($\tau$) in Fig. \ref{fig:ink_at}, it was discussed that the equilibrium crack length ($a_e$) and the work of adhesion ($W^0_\mathrm{ad}$) were determined at the equilibrium point where the crack propagation velocity reached zero. The work of adhesion and equilibrium crack length are inversely correlated, as demonstrated in equation  \eqref{eq_(7)}. This inverse relationship explains why ($a_e$) and ($W^0_\mathrm{ad}$) exhibit opposing trends with respect to $R_q$ in the presented graph. In the accompanying graph, both $a_e$ and $W^0_\mathrm{ad}$ exhibit a non-monotonic variation with respect to surface roughness ($R_q$). Specifically, within the range of grit sizes from $N=120$ to $N=320$, a decrease in $R_q$ leads to a decrease in $a_e$ and an increase in $W^0_\mathrm{ad}$. However, in the range of $N=320$ to $N=1200$, a decrease in $R_q$ results in an increase in $a_e$ and a decrease in $W^0_\mathrm{ad}$. Subsequently, for $N=1200$ to $N=2000$, as $R_q$ decreases, $a_e$ decreases while $W^0_\mathrm{ad}$ increases. Overall, within the range of grit sizes ($N$) from 120 to 2000, two optimum points are identified: at $N=320$, $W^0_\mathrm{ad}$ attains its maximum value, whereas at $N=1200$, it reaches its minimum value. The non-monotonic behavior of $a_e$ and $W^0_\mathrm{ad}$ as a function of surface roughness $R_q$ arises due to variations in the real area of contact ($A_r$) at the complementary interface of rough PDMS elastomer surfaces. The real area of contact $A_r$ is influenced by the diameter ($d$) of the asperities, the number of asperities ($n$) at the interface, and the surface roughness  ($R_q$) of the two complementary rough PDMS elastomer surfaces. As the grit size ($N$) of the sandpapers increases from $N=120$ to $N=2000$, the diameter of the asperities ($d$) and the surface roughness $R_q$ decrease, while the number of contact points or a number of asperities ($n$) at the interface increases. This inverse relationship among $d$, $R_q$, and $n$ results in two optimal points at $N=320$ and another one at $N=1200$ within the same range of grit sizes. This phenomenon, i.e., the dependence of $W^0_\mathrm{ad}$ on $A_r$, can be understood from Fig.\ref{fig:ink_Wad_Ar_N}.

Figure \ref{fig:ink_Wad_Ar_N} (A) illustrates the relationship between the work of adhesion ($W^0_\mathrm{ad}$), the real area of contact ($A_r$), and the grit sizes ($N$). The real area of contact corresponds to different asperity geometries, including hemispheroidal, conical, and a combination of both shapes, as illustrated in Fig. \ref{fig:ink_Wad_Ar_N} (B) (a), (b), and (c), respectively. These are denoted by distinct symbols: $A_{r(s)}$ for hemispheroidal asperities, $A_{r(c)}$  for conical asperities, and $A_{r(s+c)}$ for a combination of both hemispheroidal and conical asperities. From the Eqs.~\ref{eq_(2)}, \ref{eq_(3)}, and \ref{eq_(4)}, it is evident that the real area of contact in each case depends on three key parameters: the number of contact points or the number of asperities ($n$),  the diameter of the asperities ($d$), and the height or root mean square (RMS) roughness ($R_q$) of the asperities. The number of contact points and the diameter of the asperities exhibit an inverse relationship; that is, as the grit size ($N$) of the asperities increases from $N=120$ to $N=2000$, both the surface roughness ($R_q$) and the asperity diameter decrease, while the number of contact points increases. Thus, there should be optimal points within the range of $N=120$ to $N=2000$, at which the real area of contact, $A_r$, reaches its maximum and minimum values. The first such point occurs at $N=320$, where $A_r$ attains its maximum value, while another occurs at $N=1200$, where $A_r$, reaches its minimum value. Consequently, the real area of contact, $A_r$, exhibits a non-monotonic variation with $N$. Specifically, $A_r$ increases from $N=120$ to $N=320$, then decreases from $N=320$ to $N=1200$, and subsequently increases again from $N=1200$ to $N=2000$. Since the work of adhesion, ($W^0_\mathrm{ad}$), depends on the real area of contact, it follows the same non-monotonic trend as $A_r$ over the range $N=120$ to $N=2000$.

\section{Conclusions}
In this work, we have studied the effect of surface roughness on the adhesion behavior of a PDMS elastomer when two contacting PDMS elastomer surfaces form a complementary interface. Specifically, this interface is created by two surfaces that are replicas of each other, resulting in an exact peak-to-valley connection at the interface. The roughness on the PDMS elastomer surfaces was created using sandpapers of varying grit sizes, ranging from $N=120$ to $N=2000$. The surface roughness of the PDMS elastomer was measured using a stylus profilometer, and we observed that it decreased as the grit size of the sandpaper increased. Subsequently, to evaluate adhesion at the interface of two complementary rough surfaces of the PDMS elastomer, we employed the wedge test approach. Our results indicate that both the equilibrium crack length and the work of adhesion exhibit non-monotonic behavior with respect to the grit size of the sandpaper. Specifically, in the range of $N=120$ to $N=320$, the equilibrium crack length decreases while the work of adhesion increases. Between $N=320$ and $N=1200$, the equilibrium crack length increases while the work of adhesion decreases. Finally, in the range of $N=1200$ to $N=2000$, the equilibrium crack length begins to decrease again, while the work of adhesion increases. Thus, within the grit size range of $N=120$ to $N=2000$, we identified two optimal points. The first optimal point occurs at $N=320$, where the work of adhesion reaches its maximum value, while the second optimal point occurs at $N=1200$, where the work of adhesion attains its minimum value. To understand the non-monotonic variation in the work of adhesion with respect to the surface roughness of the PDMS elastomer or the grit sizes of the sandpaper, we assumed and analyzed three distinct asperity shapes present on different grit sizes of sandpaper.  These include hemispheroidal asperities, conical asperities, and a combination of both hemispheroidal and conical asperities. The diameters of these asperities were determined using high-resolution images of the rough surfaces of the PDMS elastomer, processed through ImageJ software. Using key parameters such as asperity diameter ($d$), surface roughness ($R_q$), and the number of contact points or asperities ($n$), we evaluated the real area of contact ($A_r$). Subsequently, we plotted $A_r$, the work of adhesion $W^0_\mathrm{ad}$ against the grit size ($N$) and observed that both $A_r$ and $W^0_\mathrm{ad}$ exhibited the same trend of variation with respect to $N$. Thus, both $A_r$ and $W^0_\mathrm{ad}$ increase with respect to $N$ in the range of $N=120$ to $N=320$, decrease between $N=320$ and $N=1200$, and then increase again from $N=1200$ to $N=2000$. Furthermore, within the range of $N=120$ to $N=2000$, the maximum value is observed at $N=320$, while the minimum value occurs at $N=1200$. \par

This study demonstrates various applications in the fabrication of micro- and nanoscale devices. These include triboelectric nanogenerators, microfluidic devices, and distinct electronic sensors. The proposed method for evaluating adhesion at the interface of two contacting surfaces can also be applied to assess adhesion between complementary rough surfaces of elastomers (PDMS), and various metals and alloys.

\section*{Conflicts of interest}
The authors declare no conflict of interest.

\section*{Acknowledgements}
MKS thanks the Science and Engineering Research Board (SERB), India, for the financial support provided under the Start-up Research Grant (SRG) scheme (SRG/2020/000938). KK acknowledges the funding support from DST under FIST (SR/FST/PS-II/2021/170) and AMT (DST/TDT/AM/2022/285) schemes, ANRF (CRG/2023/008576) under CRG scheme, and MoE (MoE-STARS/STARS-2/2023-0773) under STARS scheme.

\bibliography{pre}

\begin{thebibliography}{75}%
\makeatletter
\providecommand \@ifxundefined [1]{%
 \@ifx{#1\undefined}
}%
\providecommand \@ifnum [1]{%
 \ifnum #1\expandafter \@firstoftwo
 \else \expandafter \@secondoftwo
 \fi
}%
\providecommand \@ifx [1]{%
 \ifx #1\expandafter \@firstoftwo
 \else \expandafter \@secondoftwo
 \fi
}%
\providecommand \natexlab [1]{#1}%
\providecommand \enquote  [1]{``#1''}%
\providecommand \bibnamefont  [1]{#1}%
\providecommand \bibfnamefont [1]{#1}%
\providecommand \citenamefont [1]{#1}%
\providecommand \href@noop [0]{\@secondoftwo}%
\providecommand \href [0]{\begingroup \@sanitize@url \@href}%
\providecommand \@href[1]{\@@startlink{#1}\@@href}%
\providecommand \@@href[1]{\endgroup#1\@@endlink}%
\providecommand \@sanitize@url [0]{\catcode `\\12\catcode `\$12\catcode
  `\&12\catcode `\#12\catcode `\^12\catcode `\_12\catcode `\%12\relax}%
\providecommand \@@startlink[1]{}%
\providecommand \@@endlink[0]{}%
\providecommand \url  [0]{\begingroup\@sanitize@url \@url }%
\providecommand \@url [1]{\endgroup\@href {#1}{\urlprefix }}%
\providecommand \urlprefix  [0]{URL }%
\providecommand \Eprint [0]{\href }%
\providecommand \doibase [0]{http://dx.doi.org/}%
\providecommand \selectlanguage [0]{\@gobble}%
\providecommand \bibinfo  [0]{\@secondoftwo}%
\providecommand \bibfield  [0]{\@secondoftwo}%
\providecommand \translation [1]{[#1]}%
\providecommand \BibitemOpen [0]{}%
\providecommand \bibitemStop [0]{}%
\providecommand \bibitemNoStop [0]{.\EOS\space}%
\providecommand \EOS [0]{\spacefactor3000\relax}%
\providecommand \BibitemShut  [1]{\csname bibitem#1\endcsname}%
\let\auto@bib@innerbib\@empty
%</preamble>
\bibitem [{\citenamefont {Niewiarowski}\ \emph {et~al.}(2016)\citenamefont
  {Niewiarowski}, \citenamefont {Stark},\ and\ \citenamefont
  {Dhinojwala}}]{niewiarowski2016sticking}%
  \BibitemOpen
  \bibfield  {author} {\bibinfo {author} {\bibfnamefont {P.~H.}\ \bibnamefont
  {Niewiarowski}}, \bibinfo {author} {\bibfnamefont {A.~Y.}\ \bibnamefont
  {Stark}}, \ and\ \bibinfo {author} {\bibfnamefont {A.}~\bibnamefont
  {Dhinojwala}},\ }\href@noop {} {\bibfield  {journal} {\bibinfo  {journal}
  {Journal of Experimental Biology}\ }\textbf {\bibinfo {volume} {219}},\
  \bibinfo {pages} {912} (\bibinfo {year} {2016})}\BibitemShut {NoStop}%
\bibitem [{\citenamefont {Ayyildiz}\ \emph {et~al.}(2018)\citenamefont
  {Ayyildiz}, \citenamefont {Scaraggi}, \citenamefont {Sirin}, \citenamefont
  {Basdogan},\ and\ \citenamefont {Persson}}]{ayyildiz2018contact}%
  \BibitemOpen
  \bibfield  {author} {\bibinfo {author} {\bibfnamefont {M.}~\bibnamefont
  {Ayyildiz}}, \bibinfo {author} {\bibfnamefont {M.}~\bibnamefont {Scaraggi}},
  \bibinfo {author} {\bibfnamefont {O.}~\bibnamefont {Sirin}}, \bibinfo
  {author} {\bibfnamefont {C.}~\bibnamefont {Basdogan}}, \ and\ \bibinfo
  {author} {\bibfnamefont {B.~N.}\ \bibnamefont {Persson}},\ }\href@noop {}
  {\bibfield  {journal} {\bibinfo  {journal} {Proceedings of the National
  Academy of Sciences}\ }\textbf {\bibinfo {volume} {115}},\ \bibinfo {pages}
  {12668} (\bibinfo {year} {2018})}\BibitemShut {NoStop}%
\bibitem [{\citenamefont {Geim}\ \emph {et~al.}(2003)\citenamefont {Geim},
  \citenamefont {Dubonos}, \citenamefont {Grigorieva}, \citenamefont
  {Novoselov}, \citenamefont {Zhukov},\ and\ \citenamefont
  {Shapoval}}]{geim2003microfabricated}%
  \BibitemOpen
  \bibfield  {author} {\bibinfo {author} {\bibfnamefont {A.~K.}\ \bibnamefont
  {Geim}}, \bibinfo {author} {\bibfnamefont {S.}~\bibnamefont {Dubonos}},
  \bibinfo {author} {\bibfnamefont {I.~V.}\ \bibnamefont {Grigorieva}},
  \bibinfo {author} {\bibfnamefont {K.~S.}\ \bibnamefont {Novoselov}}, \bibinfo
  {author} {\bibfnamefont {A.}~\bibnamefont {Zhukov}}, \ and\ \bibinfo {author}
  {\bibfnamefont {S.~Y.}\ \bibnamefont {Shapoval}},\ }\href@noop {} {\bibfield
  {journal} {\bibinfo  {journal} {Nature materials}\ }\textbf {\bibinfo
  {volume} {2}},\ \bibinfo {pages} {461} (\bibinfo {year} {2003})}\BibitemShut
  {NoStop}%
\bibitem [{\citenamefont {Popa}\ and\ \citenamefont
  {Stephanou}(2004)}]{popa2004micro}%
  \BibitemOpen
  \bibfield  {author} {\bibinfo {author} {\bibfnamefont {D.~O.}\ \bibnamefont
  {Popa}}\ and\ \bibinfo {author} {\bibfnamefont {H.~E.}\ \bibnamefont
  {Stephanou}},\ }\href@noop {} {\bibfield  {journal} {\bibinfo  {journal}
  {Journal of manufacturing processes}\ }\textbf {\bibinfo {volume} {6}},\
  \bibinfo {pages} {52} (\bibinfo {year} {2004})}\BibitemShut {NoStop}%
\bibitem [{\citenamefont {Carlson}\ \emph {et~al.}(2012)\citenamefont
  {Carlson}, \citenamefont {Bowen}, \citenamefont {Huang}, \citenamefont
  {Nuzzo},\ and\ \citenamefont {Rogers}}]{carlson2012transfer}%
  \BibitemOpen
  \bibfield  {author} {\bibinfo {author} {\bibfnamefont {A.}~\bibnamefont
  {Carlson}}, \bibinfo {author} {\bibfnamefont {A.~M.}\ \bibnamefont {Bowen}},
  \bibinfo {author} {\bibfnamefont {Y.}~\bibnamefont {Huang}}, \bibinfo
  {author} {\bibfnamefont {R.~G.}\ \bibnamefont {Nuzzo}}, \ and\ \bibinfo
  {author} {\bibfnamefont {J.~A.}\ \bibnamefont {Rogers}},\ }\href@noop {}
  {\bibfield  {journal} {\bibinfo  {journal} {Advanced Materials}\ }\textbf
  {\bibinfo {volume} {24}},\ \bibinfo {pages} {5284} (\bibinfo {year}
  {2012})}\BibitemShut {NoStop}%
\bibitem [{\citenamefont {Ouyang}\ \emph {et~al.}(2022)\citenamefont {Ouyang},
  \citenamefont {Chen}, \citenamefont {Yan}, \citenamefont {Qin},\ and\
  \citenamefont {Liu}}]{ouyang2022mechanical}%
  \BibitemOpen
  \bibfield  {author} {\bibinfo {author} {\bibfnamefont {Z.}~\bibnamefont
  {Ouyang}}, \bibinfo {author} {\bibfnamefont {Y.}~\bibnamefont {Chen}},
  \bibinfo {author} {\bibfnamefont {Y.}~\bibnamefont {Yan}}, \bibinfo {author}
  {\bibfnamefont {H.}~\bibnamefont {Qin}}, \ and\ \bibinfo {author}
  {\bibfnamefont {Y.}~\bibnamefont {Liu}},\ }\href@noop {} {\bibfield
  {journal} {\bibinfo  {journal} {International Journal of Solids and
  Structures}\ }\textbf {\bibinfo {volume} {243}},\ \bibinfo {pages} {111589}
  (\bibinfo {year} {2022})}\BibitemShut {NoStop}%
\bibitem [{\citenamefont {Sherge}\ and\ \citenamefont
  {Gorb}(2001)}]{sherge2001biological}%
  \BibitemOpen
  \bibfield  {author} {\bibinfo {author} {\bibfnamefont {M.}~\bibnamefont
  {Sherge}}\ and\ \bibinfo {author} {\bibfnamefont {S.}~\bibnamefont {Gorb}},\
  }\href@noop {} {\enquote {\bibinfo {title} {Biological micro-and
  nano-tribology-nature’s solutions},}\ } (\bibinfo {year}
  {2001})\BibitemShut {NoStop}%
\bibitem [{\citenamefont {Gorb}(1999)}]{gorb1999evolution}%
  \BibitemOpen
  \bibfield  {author} {\bibinfo {author} {\bibfnamefont {S.~N.}\ \bibnamefont
  {Gorb}},\ }\href@noop {} {\bibfield  {journal} {\bibinfo  {journal}
  {Proceedings of the Royal Society of London. Series B: Biological Sciences}\
  }\textbf {\bibinfo {volume} {266}},\ \bibinfo {pages} {525} (\bibinfo {year}
  {1999})}\BibitemShut {NoStop}%
\bibitem [{\citenamefont {Maurya}\ \emph
  {et~al.}(2022{\natexlab{a}})\citenamefont {Maurya}, \citenamefont {Wu},
  \citenamefont {Singh},\ and\ \citenamefont {Mukherji}}]{Manoj_macrolett}%
  \BibitemOpen
  \bibfield  {author} {\bibinfo {author} {\bibfnamefont {M.~K.}\ \bibnamefont
  {Maurya}}, \bibinfo {author} {\bibfnamefont {J.}~\bibnamefont {Wu}}, \bibinfo
  {author} {\bibfnamefont {M.~K.}\ \bibnamefont {Singh}}, \ and\ \bibinfo
  {author} {\bibfnamefont {D.}~\bibnamefont {Mukherji}},\ }\href@noop {}
  {\bibfield  {journal} {\bibinfo  {journal} {ACS Macro Lett.}\ }\textbf
  {\bibinfo {volume} {11}},\ \bibinfo {pages} {925} (\bibinfo {year}
  {2022}{\natexlab{a}})}\BibitemShut {NoStop}%
\bibitem [{\citenamefont {Mukherji}\ and\ \citenamefont
  {Singh}(2021)}]{MukherjiPRM21}%
  \BibitemOpen
  \bibfield  {author} {\bibinfo {author} {\bibfnamefont {D.}~\bibnamefont
  {Mukherji}}\ and\ \bibinfo {author} {\bibfnamefont {M.~K.}\ \bibnamefont
  {Singh}},\ }\href@noop {} {\bibfield  {journal} {\bibinfo  {journal} {Phys.
  Rev. Mater.}\ }\textbf {\bibinfo {volume} {5}},\ \bibinfo {pages} {025602}
  (\bibinfo {year} {2021})}\BibitemShut {NoStop}%
\bibitem [{\citenamefont {Ain}\ \emph {et~al.}(2024{\natexlab{a}})\citenamefont
  {Ain}, \citenamefont {Wani}, \citenamefont {Sehgal},\ and\ \citenamefont
  {Singh}}]{ain2024insights}%
  \BibitemOpen
  \bibfield  {author} {\bibinfo {author} {\bibfnamefont {Q.~U.}\ \bibnamefont
  {Ain}}, \bibinfo {author} {\bibfnamefont {M.}~\bibnamefont {Wani}}, \bibinfo
  {author} {\bibfnamefont {R.}~\bibnamefont {Sehgal}}, \ and\ \bibinfo {author}
  {\bibfnamefont {M.~K.}\ \bibnamefont {Singh}},\ }\href@noop {} {\bibfield
  {journal} {\bibinfo  {journal} {Tribology International}\ }\textbf {\bibinfo
  {volume} {191}},\ \bibinfo {pages} {109174} (\bibinfo {year}
  {2024}{\natexlab{a}})}\BibitemShut {NoStop}%
\bibitem [{\citenamefont {Ain}\ \emph {et~al.}(2024{\natexlab{b}})\citenamefont
  {Ain}, \citenamefont {Wani}, \citenamefont {Sehgal},\ and\ \citenamefont
  {Singh}}]{ain2024109702}%
  \BibitemOpen
  \bibfield  {author} {\bibinfo {author} {\bibfnamefont {Q.~U.}\ \bibnamefont
  {Ain}}, \bibinfo {author} {\bibfnamefont {M.}~\bibnamefont {Wani}}, \bibinfo
  {author} {\bibfnamefont {R.}~\bibnamefont {Sehgal}}, \ and\ \bibinfo {author}
  {\bibfnamefont {M.~K.}\ \bibnamefont {Singh}},\ }\href {\doibase
  https://doi.org/10.1016/j.triboint.2024.109702} {\bibfield  {journal}
  {\bibinfo  {journal} {Tribology International}\ }\textbf {\bibinfo {volume}
  {196}},\ \bibinfo {pages} {109702} (\bibinfo {year}
  {2024}{\natexlab{b}})}\BibitemShut {NoStop}%
\bibitem [{\citenamefont {Ain}\ \emph {et~al.}(2024{\natexlab{c}})\citenamefont
  {Ain}, \citenamefont {Wani}, \citenamefont {Sehgal},\ and\ \citenamefont
  {Singh}}]{ain202433817}%
  \BibitemOpen
  \bibfield  {author} {\bibinfo {author} {\bibfnamefont {Q.~U.}\ \bibnamefont
  {Ain}}, \bibinfo {author} {\bibfnamefont {M.}~\bibnamefont {Wani}}, \bibinfo
  {author} {\bibfnamefont {R.}~\bibnamefont {Sehgal}}, \ and\ \bibinfo {author}
  {\bibfnamefont {M.~K.}\ \bibnamefont {Singh}},\ }\href {\doibase
  https://doi.org/10.1016/j.ceramint.2024.06.201} {\bibfield  {journal}
  {\bibinfo  {journal} {Ceramics International}\ }\textbf {\bibinfo {volume}
  {50}},\ \bibinfo {pages} {33817} (\bibinfo {year}
  {2024}{\natexlab{c}})}\BibitemShut {NoStop}%
\bibitem [{\citenamefont {Ain}\ \emph {et~al.}(2024{\natexlab{d}})\citenamefont
  {Ain}, \citenamefont {Wani}, \citenamefont {Sehgal},\ and\ \citenamefont
  {Singh}}]{ain2024112955}%
  \BibitemOpen
  \bibfield  {author} {\bibinfo {author} {\bibfnamefont {Q.~U.}\ \bibnamefont
  {Ain}}, \bibinfo {author} {\bibfnamefont {M.}~\bibnamefont {Wani}}, \bibinfo
  {author} {\bibfnamefont {R.}~\bibnamefont {Sehgal}}, \ and\ \bibinfo {author}
  {\bibfnamefont {M.~K.}\ \bibnamefont {Singh}},\ }\href {\doibase
  https://doi.org/10.1016/j.commatsci.2024.112955} {\bibfield  {journal}
  {\bibinfo  {journal} {Computational Materials Science}\ }\textbf {\bibinfo
  {volume} {239}},\ \bibinfo {pages} {112955} (\bibinfo {year}
  {2024}{\natexlab{d}})}\BibitemShut {NoStop}%
\bibitem [{\citenamefont {Maurya}\ \emph
  {et~al.}(2022{\natexlab{b}})\citenamefont {Maurya}, \citenamefont {Ruscher},
  \citenamefont {Mukherji},\ and\ \citenamefont
  {Singh}}]{maurya2022computational}%
  \BibitemOpen
  \bibfield  {author} {\bibinfo {author} {\bibfnamefont {M.~K.}\ \bibnamefont
  {Maurya}}, \bibinfo {author} {\bibfnamefont {C.}~\bibnamefont {Ruscher}},
  \bibinfo {author} {\bibfnamefont {D.}~\bibnamefont {Mukherji}}, \ and\
  \bibinfo {author} {\bibfnamefont {M.~K.}\ \bibnamefont {Singh}},\ }\href@noop
  {} {\bibfield  {journal} {\bibinfo  {journal} {Physical Review E}\ }\textbf
  {\bibinfo {volume} {106}},\ \bibinfo {pages} {014501} (\bibinfo {year}
  {2022}{\natexlab{b}})}\BibitemShut {NoStop}%
\bibitem [{\citenamefont {Maurya}\ and\ \citenamefont
  {Singh}(2023)}]{maurya2023computational}%
  \BibitemOpen
  \bibfield  {author} {\bibinfo {author} {\bibfnamefont {M.~K.}\ \bibnamefont
  {Maurya}}\ and\ \bibinfo {author} {\bibfnamefont {M.~K.}\ \bibnamefont
  {Singh}},\ }\href@noop {} {\bibfield  {journal} {\bibinfo  {journal}
  {International Journal of Advances in Engineering Sciences and Applied
  Mathematics}\ }\textbf {\bibinfo {volume} {15}},\ \bibinfo {pages} {196}
  (\bibinfo {year} {2023})}\BibitemShut {NoStop}%
\bibitem [{\citenamefont {Singh}\ \emph {et~al.}(2016)\citenamefont {Singh},
  \citenamefont {Ilg}, \citenamefont {Espinosa-Marzal}, \citenamefont
  {Kr{\"o}ger},\ and\ \citenamefont {Spencer}}]{singh2016effect}%
  \BibitemOpen
  \bibfield  {author} {\bibinfo {author} {\bibfnamefont {M.~K.}\ \bibnamefont
  {Singh}}, \bibinfo {author} {\bibfnamefont {P.}~\bibnamefont {Ilg}}, \bibinfo
  {author} {\bibfnamefont {R.~M.}\ \bibnamefont {Espinosa-Marzal}}, \bibinfo
  {author} {\bibfnamefont {M.}~\bibnamefont {Kr{\"o}ger}}, \ and\ \bibinfo
  {author} {\bibfnamefont {N.~D.}\ \bibnamefont {Spencer}},\ }\href@noop {}
  {\bibfield  {journal} {\bibinfo  {journal} {Tribology Letters}\ }\textbf
  {\bibinfo {volume} {63}},\ \bibinfo {pages} {1} (\bibinfo {year}
  {2016})}\BibitemShut {NoStop}%
\bibitem [{\citenamefont {Singh}\ \emph {et~al.}(2018)\citenamefont {Singh},
  \citenamefont {Kang}, \citenamefont {Ilg}, \citenamefont {Crockett},
  \citenamefont {Kröger},\ and\ \citenamefont {Spencer}}]{singh2018combined}%
  \BibitemOpen
  \bibfield  {author} {\bibinfo {author} {\bibfnamefont {M.~K.}\ \bibnamefont
  {Singh}}, \bibinfo {author} {\bibfnamefont {C.}~\bibnamefont {Kang}},
  \bibinfo {author} {\bibfnamefont {P.}~\bibnamefont {Ilg}}, \bibinfo {author}
  {\bibfnamefont {R.}~\bibnamefont {Crockett}}, \bibinfo {author}
  {\bibfnamefont {M.}~\bibnamefont {Kröger}}, \ and\ \bibinfo {author}
  {\bibfnamefont {N.~D.}\ \bibnamefont {Spencer}},\ }\href@noop {} {\bibfield
  {journal} {\bibinfo  {journal} {Macromolecules}\ }\textbf {\bibinfo {volume}
  {51}},\ \bibinfo {pages} {10174} (\bibinfo {year} {2018})}\BibitemShut
  {NoStop}%
\bibitem [{\citenamefont {Packham}(2003)}]{packham2003surface}%
  \BibitemOpen
  \bibfield  {author} {\bibinfo {author} {\bibfnamefont {D.~E.}\ \bibnamefont
  {Packham}},\ }\href@noop {} {\bibfield  {journal} {\bibinfo  {journal}
  {International journal of adhesion and adhesives}\ }\textbf {\bibinfo
  {volume} {23}},\ \bibinfo {pages} {437} (\bibinfo {year} {2003})}\BibitemShut
  {NoStop}%
\bibitem [{\citenamefont {Kersey}\ \emph {et~al.}(2009)\citenamefont {Kersey},
  \citenamefont {Ebacher}, \citenamefont {Bazargan}, \citenamefont {Wang},\
  and\ \citenamefont {Stoeber}}]{kersey2009effect}%
  \BibitemOpen
  \bibfield  {author} {\bibinfo {author} {\bibfnamefont {L.}~\bibnamefont
  {Kersey}}, \bibinfo {author} {\bibfnamefont {V.}~\bibnamefont {Ebacher}},
  \bibinfo {author} {\bibfnamefont {V.}~\bibnamefont {Bazargan}}, \bibinfo
  {author} {\bibfnamefont {R.}~\bibnamefont {Wang}}, \ and\ \bibinfo {author}
  {\bibfnamefont {B.}~\bibnamefont {Stoeber}},\ }\href@noop {} {\bibfield
  {journal} {\bibinfo  {journal} {Lab on a Chip}\ }\textbf {\bibinfo {volume}
  {9}},\ \bibinfo {pages} {1002} (\bibinfo {year} {2009})}\BibitemShut
  {NoStop}%
\bibitem [{\citenamefont {Tsai}\ \emph {et~al.}(2011)\citenamefont {Tsai},
  \citenamefont {Dahlquist}, \citenamefont {Kim},\ and\ \citenamefont
  {Nordin}}]{tsai2011bonding}%
  \BibitemOpen
  \bibfield  {author} {\bibinfo {author} {\bibfnamefont {L.-F.}\ \bibnamefont
  {Tsai}}, \bibinfo {author} {\bibfnamefont {W.~C.}\ \bibnamefont {Dahlquist}},
  \bibinfo {author} {\bibfnamefont {S.}~\bibnamefont {Kim}}, \ and\ \bibinfo
  {author} {\bibfnamefont {G.~P.}\ \bibnamefont {Nordin}},\ }\href@noop {}
  {\bibfield  {journal} {\bibinfo  {journal} {Journal of Micro/Nanolithography,
  MEMS and MOEMS}\ }\textbf {\bibinfo {volume} {10}},\ \bibinfo {pages}
  {043009} (\bibinfo {year} {2011})}\BibitemShut {NoStop}%
\bibitem [{\citenamefont {Samel}\ \emph {et~al.}(2007)\citenamefont {Samel},
  \citenamefont {Chowdhury},\ and\ \citenamefont
  {Stemme}}]{samel2007fabrication}%
  \BibitemOpen
  \bibfield  {author} {\bibinfo {author} {\bibfnamefont {B.}~\bibnamefont
  {Samel}}, \bibinfo {author} {\bibfnamefont {M.~K.}\ \bibnamefont
  {Chowdhury}}, \ and\ \bibinfo {author} {\bibfnamefont {G.}~\bibnamefont
  {Stemme}},\ }\href@noop {} {\bibfield  {journal} {\bibinfo  {journal}
  {Journal of Micromechanics and Microengineering}\ }\textbf {\bibinfo {volume}
  {17}},\ \bibinfo {pages} {1710} (\bibinfo {year} {2007})}\BibitemShut
  {NoStop}%
\bibitem [{\citenamefont {Xiong}\ \emph {et~al.}(2014)\citenamefont {Xiong},
  \citenamefont {Chen},\ and\ \citenamefont {Zhou}}]{xiong2014adhesion}%
  \BibitemOpen
  \bibfield  {author} {\bibinfo {author} {\bibfnamefont {L.}~\bibnamefont
  {Xiong}}, \bibinfo {author} {\bibfnamefont {P.}~\bibnamefont {Chen}}, \ and\
  \bibinfo {author} {\bibfnamefont {Q.}~\bibnamefont {Zhou}},\ }\href@noop {}
  {\bibfield  {journal} {\bibinfo  {journal} {Journal of Adhesion Science and
  Technology}\ }\textbf {\bibinfo {volume} {28}},\ \bibinfo {pages} {1046}
  (\bibinfo {year} {2014})}\BibitemShut {NoStop}%
\bibitem [{\citenamefont {Bhattacharya}\ \emph {et~al.}(2005)\citenamefont
  {Bhattacharya}, \citenamefont {Datta}, \citenamefont {Berg},\ and\
  \citenamefont {Gangopadhyay}}]{bhattacharya2005studies}%
  \BibitemOpen
  \bibfield  {author} {\bibinfo {author} {\bibfnamefont {S.}~\bibnamefont
  {Bhattacharya}}, \bibinfo {author} {\bibfnamefont {A.}~\bibnamefont {Datta}},
  \bibinfo {author} {\bibfnamefont {J.~M.}\ \bibnamefont {Berg}}, \ and\
  \bibinfo {author} {\bibfnamefont {S.}~\bibnamefont {Gangopadhyay}},\
  }\href@noop {} {\bibfield  {journal} {\bibinfo  {journal} {Journal of
  microelectromechanical systems}\ }\textbf {\bibinfo {volume} {14}},\ \bibinfo
  {pages} {590} (\bibinfo {year} {2005})}\BibitemShut {NoStop}%
\bibitem [{\citenamefont {Al-Ali}\ \emph {et~al.}(2023)\citenamefont {Al-Ali},
  \citenamefont {Waheed}, \citenamefont {Dawaymeh}, \citenamefont {Alamoodi},\
  and\ \citenamefont {Alazzam}}]{al2023surface}%
  \BibitemOpen
  \bibfield  {author} {\bibinfo {author} {\bibfnamefont {A.}~\bibnamefont
  {Al-Ali}}, \bibinfo {author} {\bibfnamefont {W.}~\bibnamefont {Waheed}},
  \bibinfo {author} {\bibfnamefont {F.}~\bibnamefont {Dawaymeh}}, \bibinfo
  {author} {\bibfnamefont {N.}~\bibnamefont {Alamoodi}}, \ and\ \bibinfo
  {author} {\bibfnamefont {A.}~\bibnamefont {Alazzam}},\ }\href@noop {}
  {\bibfield  {journal} {\bibinfo  {journal} {Scientific Reports}\ }\textbf
  {\bibinfo {volume} {13}},\ \bibinfo {pages} {18141} (\bibinfo {year}
  {2023})}\BibitemShut {NoStop}%
\bibitem [{\citenamefont {Sofla}\ and\ \citenamefont
  {Martin}(2010)}]{sofla2010vapor}%
  \BibitemOpen
  \bibfield  {author} {\bibinfo {author} {\bibfnamefont {A.~Y.}\ \bibnamefont
  {Sofla}}\ and\ \bibinfo {author} {\bibfnamefont {C.}~\bibnamefont {Martin}},\
  }\href@noop {} {\bibfield  {journal} {\bibinfo  {journal} {Lab on a Chip}\
  }\textbf {\bibinfo {volume} {10}},\ \bibinfo {pages} {250} (\bibinfo {year}
  {2010})}\BibitemShut {NoStop}%
\bibitem [{\citenamefont {Beh}\ \emph {et~al.}(2012)\citenamefont {Beh},
  \citenamefont {Zhou},\ and\ \citenamefont {Wang}}]{beh2012pdms}%
  \BibitemOpen
  \bibfield  {author} {\bibinfo {author} {\bibfnamefont {C.~W.}\ \bibnamefont
  {Beh}}, \bibinfo {author} {\bibfnamefont {W.}~\bibnamefont {Zhou}}, \ and\
  \bibinfo {author} {\bibfnamefont {T.-H.}\ \bibnamefont {Wang}},\ }\href@noop
  {} {\bibfield  {journal} {\bibinfo  {journal} {Lab on a Chip}\ }\textbf
  {\bibinfo {volume} {12}},\ \bibinfo {pages} {4120} (\bibinfo {year}
  {2012})}\BibitemShut {NoStop}%
\bibitem [{\citenamefont {Kumar}\ \emph
  {et~al.}(2025{\natexlab{a}})\citenamefont {Kumar}, \citenamefont {Majhi},
  \citenamefont {Khare},\ and\ \citenamefont {Singh}}]{kumar2025adhesion}%
  \BibitemOpen
  \bibfield  {author} {\bibinfo {author} {\bibfnamefont {S.}~\bibnamefont
  {Kumar}}, \bibinfo {author} {\bibfnamefont {C.}~\bibnamefont {Majhi}},
  \bibinfo {author} {\bibfnamefont {K.}~\bibnamefont {Khare}}, \ and\ \bibinfo
  {author} {\bibfnamefont {M.~K.}\ \bibnamefont {Singh}},\ }\href@noop {}
  {\bibfield  {journal} {\bibinfo  {journal} {Soft Matter}\ }\textbf {\bibinfo
  {volume} {21}},\ \bibinfo {pages} {2493} (\bibinfo {year}
  {2025}{\natexlab{a}})}\BibitemShut {NoStop}%
\bibitem [{\citenamefont {Kumar}\ \emph
  {et~al.}(2025{\natexlab{b}})\citenamefont {Kumar}, \citenamefont {Khare},\
  and\ \citenamefont {Singh}}]{kumar2025controlling}%
  \BibitemOpen
  \bibfield  {author} {\bibinfo {author} {\bibfnamefont {S.}~\bibnamefont
  {Kumar}}, \bibinfo {author} {\bibfnamefont {K.}~\bibnamefont {Khare}}, \ and\
  \bibinfo {author} {\bibfnamefont {M.~K.}\ \bibnamefont {Singh}},\ }\href@noop
  {} {\bibfield  {journal} {\bibinfo  {journal} {Tribology International}\
  }\textbf {\bibinfo {volume} {209}},\ \bibinfo {pages} {110724} (\bibinfo
  {year} {2025}{\natexlab{b}})}\BibitemShut {NoStop}%
\bibitem [{\citenamefont {Vajpayee}\ \emph {et~al.}(2011)\citenamefont
  {Vajpayee}, \citenamefont {Khare}, \citenamefont {Yang}, \citenamefont
  {Hui},\ and\ \citenamefont {Jagota}}]{vajpayee2011adhesion}%
  \BibitemOpen
  \bibfield  {author} {\bibinfo {author} {\bibfnamefont {S.}~\bibnamefont
  {Vajpayee}}, \bibinfo {author} {\bibfnamefont {K.}~\bibnamefont {Khare}},
  \bibinfo {author} {\bibfnamefont {S.}~\bibnamefont {Yang}}, \bibinfo {author}
  {\bibfnamefont {C.-Y.}\ \bibnamefont {Hui}}, \ and\ \bibinfo {author}
  {\bibfnamefont {A.}~\bibnamefont {Jagota}},\ }\href@noop {} {\bibfield
  {journal} {\bibinfo  {journal} {Advanced Functional Materials}\ }\textbf
  {\bibinfo {volume} {21}},\ \bibinfo {pages} {547} (\bibinfo {year}
  {2011})}\BibitemShut {NoStop}%
\bibitem [{\citenamefont {Kroner}\ \emph {et~al.}(2010)\citenamefont {Kroner},
  \citenamefont {Maboudian},\ and\ \citenamefont {Arzt}}]{kroner2010adhesion}%
  \BibitemOpen
  \bibfield  {author} {\bibinfo {author} {\bibfnamefont {E.}~\bibnamefont
  {Kroner}}, \bibinfo {author} {\bibfnamefont {R.}~\bibnamefont {Maboudian}}, \
  and\ \bibinfo {author} {\bibfnamefont {E.}~\bibnamefont {Arzt}},\ }\href@noop
  {} {\bibfield  {journal} {\bibinfo  {journal} {Advanced Engineering
  Materials}\ }\textbf {\bibinfo {volume} {12}},\ \bibinfo {pages} {398}
  (\bibinfo {year} {2010})}\BibitemShut {NoStop}%
\bibitem [{\citenamefont {Nair}\ \emph {et~al.}(2019)\citenamefont {Nair},
  \citenamefont {Wang},\ and\ \citenamefont {Wynne}}]{nair2019afm}%
  \BibitemOpen
  \bibfield  {author} {\bibinfo {author} {\bibfnamefont {S.~S.}\ \bibnamefont
  {Nair}}, \bibinfo {author} {\bibfnamefont {C.}~\bibnamefont {Wang}}, \ and\
  \bibinfo {author} {\bibfnamefont {K.~J.}\ \bibnamefont {Wynne}},\ }\href@noop
  {} {\bibfield  {journal} {\bibinfo  {journal} {Progress in Organic Coatings}\
  }\textbf {\bibinfo {volume} {126}},\ \bibinfo {pages} {119} (\bibinfo {year}
  {2019})}\BibitemShut {NoStop}%
\bibitem [{\citenamefont {Ghatak}\ and\ \citenamefont
  {Chaudhury}(2003)}]{ghatak2003adhesion}%
  \BibitemOpen
  \bibfield  {author} {\bibinfo {author} {\bibfnamefont {A.}~\bibnamefont
  {Ghatak}}\ and\ \bibinfo {author} {\bibfnamefont {M.~K.}\ \bibnamefont
  {Chaudhury}},\ }\href@noop {} {\bibfield  {journal} {\bibinfo  {journal}
  {Langmuir}\ }\textbf {\bibinfo {volume} {19}},\ \bibinfo {pages} {2621}
  (\bibinfo {year} {2003})}\BibitemShut {NoStop}%
\bibitem [{\citenamefont {Darby}\ \emph {et~al.}(2022)\citenamefont {Darby},
  \citenamefont {Cai}, \citenamefont {Mason},\ and\ \citenamefont
  {Pham}}]{darby2022modulus}%
  \BibitemOpen
  \bibfield  {author} {\bibinfo {author} {\bibfnamefont {D.~R.}\ \bibnamefont
  {Darby}}, \bibinfo {author} {\bibfnamefont {Z.}~\bibnamefont {Cai}}, \bibinfo
  {author} {\bibfnamefont {C.~R.}\ \bibnamefont {Mason}}, \ and\ \bibinfo
  {author} {\bibfnamefont {J.~T.}\ \bibnamefont {Pham}},\ }\href@noop {}
  {\bibfield  {journal} {\bibinfo  {journal} {Journal of Applied Polymer
  Science}\ }\textbf {\bibinfo {volume} {139}},\ \bibinfo {pages} {e52412}
  (\bibinfo {year} {2022})}\BibitemShut {NoStop}%
\bibitem [{\citenamefont {Nase}\ \emph {et~al.}(2013)\citenamefont {Nase},
  \citenamefont {Ramos}, \citenamefont {Creton},\ and\ \citenamefont
  {Lindner}}]{nase2013debonding}%
  \BibitemOpen
  \bibfield  {author} {\bibinfo {author} {\bibfnamefont {J.}~\bibnamefont
  {Nase}}, \bibinfo {author} {\bibfnamefont {O.}~\bibnamefont {Ramos}},
  \bibinfo {author} {\bibfnamefont {C.}~\bibnamefont {Creton}}, \ and\ \bibinfo
  {author} {\bibfnamefont {A.}~\bibnamefont {Lindner}},\ }\href@noop {}
  {\bibfield  {journal} {\bibinfo  {journal} {The European Physical Journal E}\
  }\textbf {\bibinfo {volume} {36}},\ \bibinfo {pages} {1} (\bibinfo {year}
  {2013})}\BibitemShut {NoStop}%
\bibitem [{\citenamefont {Murphy}\ \emph {et~al.}(2020)\citenamefont {Murphy},
  \citenamefont {Dumont}, \citenamefont {Park}, \citenamefont {Kestell},
  \citenamefont {Lee},\ and\ \citenamefont {Labouriau}}]{murphy2020tailoring}%
  \BibitemOpen
  \bibfield  {author} {\bibinfo {author} {\bibfnamefont {E.~C.}\ \bibnamefont
  {Murphy}}, \bibinfo {author} {\bibfnamefont {J.~H.}\ \bibnamefont {Dumont}},
  \bibinfo {author} {\bibfnamefont {C.~H.}\ \bibnamefont {Park}}, \bibinfo
  {author} {\bibfnamefont {G.}~\bibnamefont {Kestell}}, \bibinfo {author}
  {\bibfnamefont {K.-S.}\ \bibnamefont {Lee}}, \ and\ \bibinfo {author}
  {\bibfnamefont {A.}~\bibnamefont {Labouriau}},\ }\href@noop {} {\bibfield
  {journal} {\bibinfo  {journal} {Journal of Applied Polymer Science}\ }\textbf
  {\bibinfo {volume} {137}},\ \bibinfo {pages} {48530} (\bibinfo {year}
  {2020})}\BibitemShut {NoStop}%
\bibitem [{\citenamefont {Majhi}\ \emph {et~al.}(2023)\citenamefont {Majhi},
  \citenamefont {Bhatt}, \citenamefont {Gupta},\ and\ \citenamefont
  {Khare}}]{r41_Chiranjit}%
  \BibitemOpen
  \bibfield  {author} {\bibinfo {author} {\bibfnamefont {C.}~\bibnamefont
  {Majhi}}, \bibinfo {author} {\bibfnamefont {B.}~\bibnamefont {Bhatt}},
  \bibinfo {author} {\bibfnamefont {S.}~\bibnamefont {Gupta}}, \ and\ \bibinfo
  {author} {\bibfnamefont {K.}~\bibnamefont {Khare}},\ }\href@noop {}
  {\bibfield  {journal} {\bibinfo  {journal} {arXiv e-prints}\ ,\ \bibinfo
  {pages} {arXiv}} (\bibinfo {year} {2023})}\BibitemShut {NoStop}%
\bibitem [{\citenamefont {Luan}\ and\ \citenamefont
  {Robbins}(2005)}]{luan2005breakdown}%
  \BibitemOpen
  \bibfield  {author} {\bibinfo {author} {\bibfnamefont {B.}~\bibnamefont
  {Luan}}\ and\ \bibinfo {author} {\bibfnamefont {M.~O.}\ \bibnamefont
  {Robbins}},\ }\href@noop {} {\bibfield  {journal} {\bibinfo  {journal}
  {Nature}\ }\textbf {\bibinfo {volume} {435}},\ \bibinfo {pages} {929}
  (\bibinfo {year} {2005})}\BibitemShut {NoStop}%
\bibitem [{\citenamefont {Jacobs}\ \emph {et~al.}(2013)\citenamefont {Jacobs},
  \citenamefont {Ryan}, \citenamefont {Keating}, \citenamefont {Grierson},
  \citenamefont {Lefever}, \citenamefont {Turner}, \citenamefont {Harrison},\
  and\ \citenamefont {Carpick}}]{jacobs2013effect}%
  \BibitemOpen
  \bibfield  {author} {\bibinfo {author} {\bibfnamefont {T.~D.}\ \bibnamefont
  {Jacobs}}, \bibinfo {author} {\bibfnamefont {K.~E.}\ \bibnamefont {Ryan}},
  \bibinfo {author} {\bibfnamefont {P.~L.}\ \bibnamefont {Keating}}, \bibinfo
  {author} {\bibfnamefont {D.~S.}\ \bibnamefont {Grierson}}, \bibinfo {author}
  {\bibfnamefont {J.~A.}\ \bibnamefont {Lefever}}, \bibinfo {author}
  {\bibfnamefont {K.~T.}\ \bibnamefont {Turner}}, \bibinfo {author}
  {\bibfnamefont {J.~A.}\ \bibnamefont {Harrison}}, \ and\ \bibinfo {author}
  {\bibfnamefont {R.~W.}\ \bibnamefont {Carpick}},\ }\href@noop {} {\bibfield
  {journal} {\bibinfo  {journal} {Tribology Letters}\ }\textbf {\bibinfo
  {volume} {50}},\ \bibinfo {pages} {81} (\bibinfo {year} {2013})}\BibitemShut
  {NoStop}%
\bibitem [{\citenamefont {Pastewka}\ and\ \citenamefont
  {Robbins}(2014)}]{pastewka2014contact}%
  \BibitemOpen
  \bibfield  {author} {\bibinfo {author} {\bibfnamefont {L.}~\bibnamefont
  {Pastewka}}\ and\ \bibinfo {author} {\bibfnamefont {M.~O.}\ \bibnamefont
  {Robbins}},\ }\href@noop {} {\bibfield  {journal} {\bibinfo  {journal}
  {Proceedings of the National Academy of Sciences}\ }\textbf {\bibinfo
  {volume} {111}},\ \bibinfo {pages} {3298} (\bibinfo {year}
  {2014})}\BibitemShut {NoStop}%
\bibitem [{\citenamefont {Zhang}\ \emph {et~al.}(2018)\citenamefont {Zhang},
  \citenamefont {Li}, \citenamefont {Wang}, \citenamefont {Tian}, \citenamefont
  {Liu}, \citenamefont {Ye}, \citenamefont {Liu}, \citenamefont {Zhang},\ and\
  \citenamefont {Han}}]{zhang2018high}%
  \BibitemOpen
  \bibfield  {author} {\bibinfo {author} {\bibfnamefont {X.-W.}\ \bibnamefont
  {Zhang}}, \bibinfo {author} {\bibfnamefont {G.-Z.}\ \bibnamefont {Li}},
  \bibinfo {author} {\bibfnamefont {G.-G.}\ \bibnamefont {Wang}}, \bibinfo
  {author} {\bibfnamefont {J.-L.}\ \bibnamefont {Tian}}, \bibinfo {author}
  {\bibfnamefont {Y.-L.}\ \bibnamefont {Liu}}, \bibinfo {author} {\bibfnamefont
  {D.-M.}\ \bibnamefont {Ye}}, \bibinfo {author} {\bibfnamefont
  {Z.}~\bibnamefont {Liu}}, \bibinfo {author} {\bibfnamefont {H.-Y.}\
  \bibnamefont {Zhang}}, \ and\ \bibinfo {author} {\bibfnamefont {J.-C.}\
  \bibnamefont {Han}},\ }\href@noop {} {\bibfield  {journal} {\bibinfo
  {journal} {ACS Sustainable Chemistry \& Engineering}\ }\textbf {\bibinfo
  {volume} {6}},\ \bibinfo {pages} {2283} (\bibinfo {year} {2018})}\BibitemShut
  {NoStop}%
\bibitem [{\citenamefont {Kim}\ \emph {et~al.}(2017)\citenamefont {Kim},
  \citenamefont {Lee},\ and\ \citenamefont {Choi}}]{kim2017large}%
  \BibitemOpen
  \bibfield  {author} {\bibinfo {author} {\bibfnamefont {D.}~\bibnamefont
  {Kim}}, \bibinfo {author} {\bibfnamefont {H.~M.}\ \bibnamefont {Lee}}, \ and\
  \bibinfo {author} {\bibfnamefont {Y.-K.}\ \bibnamefont {Choi}},\ }\href@noop
  {} {\bibfield  {journal} {\bibinfo  {journal} {Rsc Advances}\ }\textbf
  {\bibinfo {volume} {7}},\ \bibinfo {pages} {137} (\bibinfo {year}
  {2017})}\BibitemShut {NoStop}%
\bibitem [{\citenamefont {Rasel}\ and\ \citenamefont
  {Park}(2017)}]{rasel2017sandpaper}%
  \BibitemOpen
  \bibfield  {author} {\bibinfo {author} {\bibfnamefont {M.~S.~U.}\
  \bibnamefont {Rasel}}\ and\ \bibinfo {author} {\bibfnamefont {J.-Y.}\
  \bibnamefont {Park}},\ }\href@noop {} {\bibfield  {journal} {\bibinfo
  {journal} {Applied energy}\ }\textbf {\bibinfo {volume} {206}},\ \bibinfo
  {pages} {150} (\bibinfo {year} {2017})}\BibitemShut {NoStop}%
\bibitem [{\citenamefont {Liu}\ \emph {et~al.}(2019)\citenamefont {Liu},
  \citenamefont {Gu}, \citenamefont {Jia}, \citenamefont {Liu}, \citenamefont
  {Zhang}, \citenamefont {Wang}, \citenamefont {Zhang}, \citenamefont {Zhang},\
  and\ \citenamefont {Zhang}}]{liu2019design}%
  \BibitemOpen
  \bibfield  {author} {\bibinfo {author} {\bibfnamefont {Y.}~\bibnamefont
  {Liu}}, \bibinfo {author} {\bibfnamefont {H.}~\bibnamefont {Gu}}, \bibinfo
  {author} {\bibfnamefont {Y.}~\bibnamefont {Jia}}, \bibinfo {author}
  {\bibfnamefont {J.}~\bibnamefont {Liu}}, \bibinfo {author} {\bibfnamefont
  {H.}~\bibnamefont {Zhang}}, \bibinfo {author} {\bibfnamefont
  {R.}~\bibnamefont {Wang}}, \bibinfo {author} {\bibfnamefont {B.}~\bibnamefont
  {Zhang}}, \bibinfo {author} {\bibfnamefont {H.}~\bibnamefont {Zhang}}, \ and\
  \bibinfo {author} {\bibfnamefont {Q.}~\bibnamefont {Zhang}},\ }\href@noop {}
  {\bibfield  {journal} {\bibinfo  {journal} {Chemical Engineering Journal}\
  }\textbf {\bibinfo {volume} {356}},\ \bibinfo {pages} {318} (\bibinfo {year}
  {2019})}\BibitemShut {NoStop}%
\bibitem [{\citenamefont {Yu}\ and\ \citenamefont
  {Cheng}(2018)}]{yu2018tunable}%
  \BibitemOpen
  \bibfield  {author} {\bibinfo {author} {\bibfnamefont {Z.}~\bibnamefont
  {Yu}}\ and\ \bibinfo {author} {\bibfnamefont {H.}~\bibnamefont {Cheng}},\
  }\href@noop {} {\bibfield  {journal} {\bibinfo  {journal} {Micromachines}\
  }\textbf {\bibinfo {volume} {9}},\ \bibinfo {pages} {529} (\bibinfo {year}
  {2018})}\BibitemShut {NoStop}%
\bibitem [{\citenamefont {Hill}\ \emph {et~al.}(2016)\citenamefont {Hill},
  \citenamefont {Qian}, \citenamefont {Chen},\ and\ \citenamefont
  {Fu}}]{hill2016surface}%
  \BibitemOpen
  \bibfield  {author} {\bibinfo {author} {\bibfnamefont {S.}~\bibnamefont
  {Hill}}, \bibinfo {author} {\bibfnamefont {W.}~\bibnamefont {Qian}}, \bibinfo
  {author} {\bibfnamefont {W.}~\bibnamefont {Chen}}, \ and\ \bibinfo {author}
  {\bibfnamefont {J.}~\bibnamefont {Fu}},\ }\href@noop {} {\bibfield  {journal}
  {\bibinfo  {journal} {Biomicrofluidics}\ }\textbf {\bibinfo {volume} {10}},\
  \bibinfo {pages} {054114} (\bibinfo {year} {2016})}\BibitemShut {NoStop}%
\bibitem [{\citenamefont {Zhou}\ \emph {et~al.}(2024)\citenamefont {Zhou},
  \citenamefont {Zhai}, \citenamefont {Wang}, \citenamefont {Wang},
  \citenamefont {Zheng}, \citenamefont {Wu}, \citenamefont {Yan},\ and\
  \citenamefont {Liu}}]{zhou2024laser}%
  \BibitemOpen
  \bibfield  {author} {\bibinfo {author} {\bibfnamefont {X.}~\bibnamefont
  {Zhou}}, \bibinfo {author} {\bibfnamefont {Z.}~\bibnamefont {Zhai}}, \bibinfo
  {author} {\bibfnamefont {J.}~\bibnamefont {Wang}}, \bibinfo {author}
  {\bibfnamefont {T.}~\bibnamefont {Wang}}, \bibinfo {author} {\bibfnamefont
  {H.}~\bibnamefont {Zheng}}, \bibinfo {author} {\bibfnamefont
  {Y.}~\bibnamefont {Wu}}, \bibinfo {author} {\bibfnamefont {C.}~\bibnamefont
  {Yan}}, \ and\ \bibinfo {author} {\bibfnamefont {M.}~\bibnamefont {Liu}},\
  }\href@noop {} {\bibfield  {journal} {\bibinfo  {journal} {ACS Applied
  Polymer Materials}\ }\textbf {\bibinfo {volume} {6}},\ \bibinfo {pages}
  {7137} (\bibinfo {year} {2024})}\BibitemShut {NoStop}%
\bibitem [{\citenamefont {Liu}\ \emph {et~al.}(2023)\citenamefont {Liu},
  \citenamefont {Oyunbaatar}, \citenamefont {Shanmugasundaram}, \citenamefont
  {Kim}, \citenamefont {Lee},\ and\ \citenamefont {Lee}}]{liu2023nano}%
  \BibitemOpen
  \bibfield  {author} {\bibinfo {author} {\bibfnamefont {Y.}~\bibnamefont
  {Liu}}, \bibinfo {author} {\bibfnamefont {N.-E.}\ \bibnamefont {Oyunbaatar}},
  \bibinfo {author} {\bibfnamefont {A.}~\bibnamefont {Shanmugasundaram}},
  \bibinfo {author} {\bibfnamefont {E.-S.}\ \bibnamefont {Kim}}, \bibinfo
  {author} {\bibfnamefont {B.-K.}\ \bibnamefont {Lee}}, \ and\ \bibinfo
  {author} {\bibfnamefont {D.-W.}\ \bibnamefont {Lee}},\ }\href@noop {}
  {\bibfield  {journal} {\bibinfo  {journal} {Sensors and Actuators B:
  Chemical}\ }\textbf {\bibinfo {volume} {390}},\ \bibinfo {pages} {134014}
  (\bibinfo {year} {2023})}\BibitemShut {NoStop}%
\bibitem [{\citenamefont {Yi}\ \emph {et~al.}(2020)\citenamefont {Yi},
  \citenamefont {Ma}, \citenamefont {Mizuno}, \citenamefont {Makita},
  \citenamefont {Sugaya}, \citenamefont {Takato}, \citenamefont {Mehrvarz},
  \citenamefont {Bremner},\ and\ \citenamefont
  {Ho-Baillie}}]{yi2020application}%
  \BibitemOpen
  \bibfield  {author} {\bibinfo {author} {\bibfnamefont {C.}~\bibnamefont
  {Yi}}, \bibinfo {author} {\bibfnamefont {F.-J.}\ \bibnamefont {Ma}}, \bibinfo
  {author} {\bibfnamefont {H.}~\bibnamefont {Mizuno}}, \bibinfo {author}
  {\bibfnamefont {K.}~\bibnamefont {Makita}}, \bibinfo {author} {\bibfnamefont
  {T.}~\bibnamefont {Sugaya}}, \bibinfo {author} {\bibfnamefont
  {H.}~\bibnamefont {Takato}}, \bibinfo {author} {\bibfnamefont
  {H.}~\bibnamefont {Mehrvarz}}, \bibinfo {author} {\bibfnamefont
  {S.}~\bibnamefont {Bremner}}, \ and\ \bibinfo {author} {\bibfnamefont
  {A.}~\bibnamefont {Ho-Baillie}},\ }\href@noop {} {\bibfield  {journal}
  {\bibinfo  {journal} {Optics Express}\ }\textbf {\bibinfo {volume} {28}},\
  \bibinfo {pages} {3895} (\bibinfo {year} {2020})}\BibitemShut {NoStop}%
\bibitem [{\citenamefont {Ren}\ \emph {et~al.}(2020)\citenamefont {Ren},
  \citenamefont {Liu}, \citenamefont {Xu}, \citenamefont {Zhang}, \citenamefont
  {Shao},\ and\ \citenamefont {He}}]{ren2020high}%
  \BibitemOpen
  \bibfield  {author} {\bibinfo {author} {\bibfnamefont {L.-F.}\ \bibnamefont
  {Ren}}, \bibinfo {author} {\bibfnamefont {C.}~\bibnamefont {Liu}}, \bibinfo
  {author} {\bibfnamefont {Y.}~\bibnamefont {Xu}}, \bibinfo {author}
  {\bibfnamefont {X.}~\bibnamefont {Zhang}}, \bibinfo {author} {\bibfnamefont
  {J.}~\bibnamefont {Shao}}, \ and\ \bibinfo {author} {\bibfnamefont
  {Y.}~\bibnamefont {He}},\ }\href@noop {} {\bibfield  {journal} {\bibinfo
  {journal} {Journal of Membrane Science}\ }\textbf {\bibinfo {volume} {597}},\
  \bibinfo {pages} {117624} (\bibinfo {year} {2020})}\BibitemShut {NoStop}%
\bibitem [{\citenamefont {Brounstein}\ \emph {et~al.}(2021)\citenamefont
  {Brounstein}, \citenamefont {Zhao}, \citenamefont {Geller}, \citenamefont
  {Gupta},\ and\ \citenamefont {Labouriau}}]{brounstein2021long}%
  \BibitemOpen
  \bibfield  {author} {\bibinfo {author} {\bibfnamefont {Z.}~\bibnamefont
  {Brounstein}}, \bibinfo {author} {\bibfnamefont {J.}~\bibnamefont {Zhao}},
  \bibinfo {author} {\bibfnamefont {D.}~\bibnamefont {Geller}}, \bibinfo
  {author} {\bibfnamefont {N.}~\bibnamefont {Gupta}}, \ and\ \bibinfo {author}
  {\bibfnamefont {A.}~\bibnamefont {Labouriau}},\ }\href@noop {} {\bibfield
  {journal} {\bibinfo  {journal} {Polymers}\ }\textbf {\bibinfo {volume}
  {13}},\ \bibinfo {pages} {3125} (\bibinfo {year} {2021})}\BibitemShut
  {NoStop}%
\bibitem [{\citenamefont {An}\ \emph {et~al.}(2017)\citenamefont {An},
  \citenamefont {Guo}, \citenamefont {Lee}, \citenamefont {Jeong},
  \citenamefont {Zhao}, \citenamefont {Wang},\ and\ \citenamefont
  {Leiknes}}]{an2017pdms}%
  \BibitemOpen
  \bibfield  {author} {\bibinfo {author} {\bibfnamefont {A.~K.}\ \bibnamefont
  {An}}, \bibinfo {author} {\bibfnamefont {J.}~\bibnamefont {Guo}}, \bibinfo
  {author} {\bibfnamefont {E.-J.}\ \bibnamefont {Lee}}, \bibinfo {author}
  {\bibfnamefont {S.}~\bibnamefont {Jeong}}, \bibinfo {author} {\bibfnamefont
  {Y.}~\bibnamefont {Zhao}}, \bibinfo {author} {\bibfnamefont {Z.}~\bibnamefont
  {Wang}}, \ and\ \bibinfo {author} {\bibfnamefont {T.}~\bibnamefont
  {Leiknes}},\ }\href@noop {} {\bibfield  {journal} {\bibinfo  {journal}
  {Journal of Membrane Science}\ }\textbf {\bibinfo {volume} {525}},\ \bibinfo
  {pages} {57} (\bibinfo {year} {2017})}\BibitemShut {NoStop}%
\bibitem [{\citenamefont {Wolf}\ \emph {et~al.}(2018)\citenamefont {Wolf},
  \citenamefont {Salieb-Beugelaar},\ and\ \citenamefont
  {Hunziker}}]{wolf2018pdms}%
  \BibitemOpen
  \bibfield  {author} {\bibinfo {author} {\bibfnamefont {M.~P.}\ \bibnamefont
  {Wolf}}, \bibinfo {author} {\bibfnamefont {G.~B.}\ \bibnamefont
  {Salieb-Beugelaar}}, \ and\ \bibinfo {author} {\bibfnamefont
  {P.}~\bibnamefont {Hunziker}},\ }\href@noop {} {\bibfield  {journal}
  {\bibinfo  {journal} {Progress in Polymer Science}\ }\textbf {\bibinfo
  {volume} {83}},\ \bibinfo {pages} {97} (\bibinfo {year} {2018})}\BibitemShut
  {NoStop}%
\bibitem [{\citenamefont {Van~Poll}\ \emph {et~al.}(2007)\citenamefont
  {Van~Poll}, \citenamefont {Zhou}, \citenamefont {Ramstedt}, \citenamefont
  {Hu},\ and\ \citenamefont {Huck}}]{pr8_Micropatterning}%
  \BibitemOpen
  \bibfield  {author} {\bibinfo {author} {\bibfnamefont {M.~L.}\ \bibnamefont
  {Van~Poll}}, \bibinfo {author} {\bibfnamefont {F.}~\bibnamefont {Zhou}},
  \bibinfo {author} {\bibfnamefont {M.}~\bibnamefont {Ramstedt}}, \bibinfo
  {author} {\bibfnamefont {L.}~\bibnamefont {Hu}}, \ and\ \bibinfo {author}
  {\bibfnamefont {W.~T.}\ \bibnamefont {Huck}},\ }\href@noop {} {\bibfield
  {journal} {\bibinfo  {journal} {Angew. Chem.}\ }\textbf {\bibinfo {volume}
  {119}},\ \bibinfo {pages} {6754} (\bibinfo {year} {2007})}\BibitemShut
  {NoStop}%
\bibitem [{\citenamefont {Lopera}\ and\ \citenamefont
  {Mansano}(2012)}]{pr5_surfacemodification}%
  \BibitemOpen
  \bibfield  {author} {\bibinfo {author} {\bibfnamefont {S.}~\bibnamefont
  {Lopera}}\ and\ \bibinfo {author} {\bibfnamefont {R.}~\bibnamefont
  {Mansano}},\ }\href@noop {} {\bibfield  {journal} {\bibinfo  {journal}
  {International Scholarly Research Notices}\ }\textbf {\bibinfo {volume}
  {2012}},\ \bibinfo {pages} {5} (\bibinfo {year} {2012})}\BibitemShut
  {NoStop}%
\bibitem [{\citenamefont {Souza}\ \emph {et~al.}(2020)\citenamefont {Souza},
  \citenamefont {Marques}, \citenamefont {Balsa},\ and\ \citenamefont
  {Ribeiro}}]{souza2020characterization}%
  \BibitemOpen
  \bibfield  {author} {\bibinfo {author} {\bibfnamefont {A.}~\bibnamefont
  {Souza}}, \bibinfo {author} {\bibfnamefont {E.}~\bibnamefont {Marques}},
  \bibinfo {author} {\bibfnamefont {C.}~\bibnamefont {Balsa}}, \ and\ \bibinfo
  {author} {\bibfnamefont {J.}~\bibnamefont {Ribeiro}},\ }\href@noop {}
  {\bibfield  {journal} {\bibinfo  {journal} {Applied Sciences}\ }\textbf
  {\bibinfo {volume} {10}},\ \bibinfo {pages} {3322} (\bibinfo {year}
  {2020})}\BibitemShut {NoStop}%
\bibitem [{\citenamefont {Kuddannaya}\ \emph {et~al.}(2015)\citenamefont
  {Kuddannaya}, \citenamefont {Bao},\ and\ \citenamefont
  {Zhang}}]{pr11_biocompatibility}%
  \BibitemOpen
  \bibfield  {author} {\bibinfo {author} {\bibfnamefont {S.}~\bibnamefont
  {Kuddannaya}}, \bibinfo {author} {\bibfnamefont {J.}~\bibnamefont {Bao}}, \
  and\ \bibinfo {author} {\bibfnamefont {Y.}~\bibnamefont {Zhang}},\
  }\href@noop {} {\bibfield  {journal} {\bibinfo  {journal} {ACS Appl. Mater.
  \& Interfaces}\ }\textbf {\bibinfo {volume} {7}},\ \bibinfo {pages} {25529}
  (\bibinfo {year} {2015})}\BibitemShut {NoStop}%
\bibitem [{\citenamefont {Eduok}\ \emph {et~al.}(2017)\citenamefont {Eduok},
  \citenamefont {Faye},\ and\ \citenamefont {Szpunar}}]{eduok2017recent}%
  \BibitemOpen
  \bibfield  {author} {\bibinfo {author} {\bibfnamefont {U.}~\bibnamefont
  {Eduok}}, \bibinfo {author} {\bibfnamefont {O.}~\bibnamefont {Faye}}, \ and\
  \bibinfo {author} {\bibfnamefont {J.}~\bibnamefont {Szpunar}},\ }\href@noop
  {} {\bibfield  {journal} {\bibinfo  {journal} {Progress in Organic Coatings}\
  }\textbf {\bibinfo {volume} {111}},\ \bibinfo {pages} {124} (\bibinfo {year}
  {2017})}\BibitemShut {NoStop}%
\bibitem [{\citenamefont {Passot}\ and\ \citenamefont
  {Cabodevila}(2011)}]{passot2011mechanical}%
  \BibitemOpen
  \bibfield  {author} {\bibinfo {author} {\bibfnamefont {A.}~\bibnamefont
  {Passot}}\ and\ \bibinfo {author} {\bibfnamefont {G.}~\bibnamefont
  {Cabodevila}},\ }in\ \href@noop {} {\emph {\bibinfo {booktitle}
  {Bioelectronics, Biomedical, and Bioinspired Systems V; and Nanotechnology
  V}}},\ Vol.\ \bibinfo {volume} {8068}\ (\bibinfo {organization} {SPIE},\
  \bibinfo {year} {2011})\ pp.\ \bibinfo {pages} {113--124}\BibitemShut
  {NoStop}%
\bibitem [{\citenamefont {Gao}\ \emph {et~al.}(2020)\citenamefont {Gao},
  \citenamefont {Song}, \citenamefont {Zhang}, \citenamefont {Li},
  \citenamefont {Yang}, \citenamefont {Wang}, \citenamefont {Li}, \citenamefont
  {Zhang}, \citenamefont {Guo},\ and\ \citenamefont {Fu}}]{gao2020facile}%
  \BibitemOpen
  \bibfield  {author} {\bibinfo {author} {\bibfnamefont {Z.}~\bibnamefont
  {Gao}}, \bibinfo {author} {\bibfnamefont {G.}~\bibnamefont {Song}}, \bibinfo
  {author} {\bibfnamefont {X.}~\bibnamefont {Zhang}}, \bibinfo {author}
  {\bibfnamefont {Q.}~\bibnamefont {Li}}, \bibinfo {author} {\bibfnamefont
  {S.}~\bibnamefont {Yang}}, \bibinfo {author} {\bibfnamefont {T.}~\bibnamefont
  {Wang}}, \bibinfo {author} {\bibfnamefont {Y.}~\bibnamefont {Li}}, \bibinfo
  {author} {\bibfnamefont {L.}~\bibnamefont {Zhang}}, \bibinfo {author}
  {\bibfnamefont {L.}~\bibnamefont {Guo}}, \ and\ \bibinfo {author}
  {\bibfnamefont {Y.}~\bibnamefont {Fu}},\ }\href@noop {} {\bibfield  {journal}
  {\bibinfo  {journal} {Sensors and Actuators B: Chemical}\ }\textbf {\bibinfo
  {volume} {325}},\ \bibinfo {pages} {128810} (\bibinfo {year}
  {2020})}\BibitemShut {NoStop}%
\bibitem [{\citenamefont {Lee}\ \emph {et~al.}(2016)\citenamefont {Lee},
  \citenamefont {Kim}, \citenamefont {Han}, \citenamefont {Kim}, \citenamefont
  {Park}, \citenamefont {Jeong}, \citenamefont {Kim},\ and\ \citenamefont
  {Kim}}]{lee2016fabrication}%
  \BibitemOpen
  \bibfield  {author} {\bibinfo {author} {\bibfnamefont {J.~H.}\ \bibnamefont
  {Lee}}, \bibinfo {author} {\bibfnamefont {D.~H.}\ \bibnamefont {Kim}},
  \bibinfo {author} {\bibfnamefont {S.~W.}\ \bibnamefont {Han}}, \bibinfo
  {author} {\bibfnamefont {B.~R.}\ \bibnamefont {Kim}}, \bibinfo {author}
  {\bibfnamefont {E.~J.}\ \bibnamefont {Park}}, \bibinfo {author}
  {\bibfnamefont {M.-G.}\ \bibnamefont {Jeong}}, \bibinfo {author}
  {\bibfnamefont {J.~H.}\ \bibnamefont {Kim}}, \ and\ \bibinfo {author}
  {\bibfnamefont {Y.~D.}\ \bibnamefont {Kim}},\ }\href@noop {} {\bibfield
  {journal} {\bibinfo  {journal} {Chemical Engineering Journal}\ }\textbf
  {\bibinfo {volume} {289}},\ \bibinfo {pages} {1} (\bibinfo {year}
  {2016})}\BibitemShut {NoStop}%
\bibitem [{\citenamefont {He}\ \emph {et~al.}(2018)\citenamefont {He},
  \citenamefont {Wang}, \citenamefont {Li}, \citenamefont {Chen},\ and\
  \citenamefont {Li}}]{he2018fabrication}%
  \BibitemOpen
  \bibfield  {author} {\bibinfo {author} {\bibfnamefont {X.}~\bibnamefont
  {He}}, \bibinfo {author} {\bibfnamefont {T.}~\bibnamefont {Wang}}, \bibinfo
  {author} {\bibfnamefont {Y.}~\bibnamefont {Li}}, \bibinfo {author}
  {\bibfnamefont {J.}~\bibnamefont {Chen}}, \ and\ \bibinfo {author}
  {\bibfnamefont {J.}~\bibnamefont {Li}},\ }\href@noop {} {\bibfield  {journal}
  {\bibinfo  {journal} {Journal of membrane science}\ }\textbf {\bibinfo
  {volume} {563}},\ \bibinfo {pages} {447} (\bibinfo {year}
  {2018})}\BibitemShut {NoStop}%
\bibitem [{\citenamefont {Nguyen}\ \emph {et~al.}(2022)\citenamefont {Nguyen},
  \citenamefont {Sarkar}, \citenamefont {Tran}, \citenamefont {Moinuddin},
  \citenamefont {Saha},\ and\ \citenamefont {Ahsan}}]{nguyen2022multilayer}%
  \BibitemOpen
  \bibfield  {author} {\bibinfo {author} {\bibfnamefont {T.}~\bibnamefont
  {Nguyen}}, \bibinfo {author} {\bibfnamefont {T.}~\bibnamefont {Sarkar}},
  \bibinfo {author} {\bibfnamefont {T.}~\bibnamefont {Tran}}, \bibinfo {author}
  {\bibfnamefont {S.~M.}\ \bibnamefont {Moinuddin}}, \bibinfo {author}
  {\bibfnamefont {D.}~\bibnamefont {Saha}}, \ and\ \bibinfo {author}
  {\bibfnamefont {F.}~\bibnamefont {Ahsan}},\ }\href@noop {} {\bibfield
  {journal} {\bibinfo  {journal} {Micromachines}\ }\textbf {\bibinfo {volume}
  {13}},\ \bibinfo {pages} {1357} (\bibinfo {year} {2022})}\BibitemShut
  {NoStop}%
\bibitem [{\citenamefont {Chen}\ \emph {et~al.}(2012)\citenamefont {Chen},
  \citenamefont {Lam},\ and\ \citenamefont {Fu}}]{chen2012photolithographic}%
  \BibitemOpen
  \bibfield  {author} {\bibinfo {author} {\bibfnamefont {W.}~\bibnamefont
  {Chen}}, \bibinfo {author} {\bibfnamefont {R.~H.}\ \bibnamefont {Lam}}, \
  and\ \bibinfo {author} {\bibfnamefont {J.}~\bibnamefont {Fu}},\ }\href@noop
  {} {\bibfield  {journal} {\bibinfo  {journal} {Lab on a Chip}\ }\textbf
  {\bibinfo {volume} {12}},\ \bibinfo {pages} {391} (\bibinfo {year}
  {2012})}\BibitemShut {NoStop}%
\bibitem [{\citenamefont {Felix}\ \emph {et~al.}(2014)\citenamefont {Felix},
  \citenamefont {Santiago-Alvarado}, \citenamefont {Iturbide-Jim{\'e}nez},\
  and\ \citenamefont {Licona-Mor{\'a}n}}]{felix2014physical}%
  \BibitemOpen
  \bibfield  {author} {\bibinfo {author} {\bibfnamefont {A.~C.}\ \bibnamefont
  {Felix}}, \bibinfo {author} {\bibfnamefont {A.}~\bibnamefont
  {Santiago-Alvarado}}, \bibinfo {author} {\bibfnamefont {F.}~\bibnamefont
  {Iturbide-Jim{\'e}nez}}, \ and\ \bibinfo {author} {\bibfnamefont
  {B.}~\bibnamefont {Licona-Mor{\'a}n}},\ }\href@noop {} {\bibfield  {journal}
  {\bibinfo  {journal} {Intl. J. Eng. Sci. Innov. Tech}\ }\textbf {\bibinfo
  {volume} {3}},\ \bibinfo {pages} {563} (\bibinfo {year} {2014})}\BibitemShut
  {NoStop}%
\bibitem [{\citenamefont {Lai}\ \emph {et~al.}(2019)\citenamefont {Lai},
  \citenamefont {Altemose}, \citenamefont {White},\ and\ \citenamefont
  {Streets}}]{lai2019ratio}%
  \BibitemOpen
  \bibfield  {author} {\bibinfo {author} {\bibfnamefont {A.}~\bibnamefont
  {Lai}}, \bibinfo {author} {\bibfnamefont {N.}~\bibnamefont {Altemose}},
  \bibinfo {author} {\bibfnamefont {J.~A.}\ \bibnamefont {White}}, \ and\
  \bibinfo {author} {\bibfnamefont {A.~M.}\ \bibnamefont {Streets}},\
  }\href@noop {} {\bibfield  {journal} {\bibinfo  {journal} {Journal of
  Micromechanics and Microengineering}\ }\textbf {\bibinfo {volume} {29}},\
  \bibinfo {pages} {107001} (\bibinfo {year} {2019})}\BibitemShut {NoStop}%
\bibitem [{\citenamefont {Wiranata}\ and\ \citenamefont
  {Maeda}(2020)}]{wiranata2020implementation}%
  \BibitemOpen
  \bibfield  {author} {\bibinfo {author} {\bibfnamefont {A.}~\bibnamefont
  {Wiranata}}\ and\ \bibinfo {author} {\bibfnamefont {S.}~\bibnamefont
  {Maeda}},\ }\href@noop {} {\bibfield  {journal} {\bibinfo  {journal} {SEATUC
  journal of science and engineering}\ }\textbf {\bibinfo {volume} {1}},\
  \bibinfo {pages} {14} (\bibinfo {year} {2020})}\BibitemShut {NoStop}%
\bibitem [{\citenamefont {Li}\ \emph {et~al.}(2021)\citenamefont {Li},
  \citenamefont {Xiong}, \citenamefont {Cao}, \citenamefont {Zhu},
  \citenamefont {Lin}, \citenamefont {Zhang}, \citenamefont {Liu},
  \citenamefont {Yang}, \citenamefont {Cao},\ and\ \citenamefont
  {Chen}}]{li2021flexible}%
  \BibitemOpen
  \bibfield  {author} {\bibinfo {author} {\bibfnamefont {Y.}~\bibnamefont
  {Li}}, \bibinfo {author} {\bibfnamefont {Y.}~\bibnamefont {Xiong}}, \bibinfo
  {author} {\bibfnamefont {W.}~\bibnamefont {Cao}}, \bibinfo {author}
  {\bibfnamefont {Q.}~\bibnamefont {Zhu}}, \bibinfo {author} {\bibfnamefont
  {Y.}~\bibnamefont {Lin}}, \bibinfo {author} {\bibfnamefont {Y.}~\bibnamefont
  {Zhang}}, \bibinfo {author} {\bibfnamefont {M.}~\bibnamefont {Liu}}, \bibinfo
  {author} {\bibfnamefont {F.}~\bibnamefont {Yang}}, \bibinfo {author}
  {\bibfnamefont {K.}~\bibnamefont {Cao}}, \ and\ \bibinfo {author}
  {\bibfnamefont {R.}~\bibnamefont {Chen}},\ }\href@noop {} {\bibfield
  {journal} {\bibinfo  {journal} {Advanced Materials Interfaces}\ }\textbf
  {\bibinfo {volume} {8}},\ \bibinfo {pages} {2100872} (\bibinfo {year}
  {2021})}\BibitemShut {NoStop}%
\bibitem [{\citenamefont {Wienzek}\ and\ \citenamefont
  {Seibel}(2019)}]{wienzek2019elastomeric}%
  \BibitemOpen
  \bibfield  {author} {\bibinfo {author} {\bibfnamefont {T.}~\bibnamefont
  {Wienzek}}\ and\ \bibinfo {author} {\bibfnamefont {A.}~\bibnamefont
  {Seibel}},\ }\href@noop {} {\bibfield  {journal} {\bibinfo  {journal}
  {Advanced engineering materials}\ }\textbf {\bibinfo {volume} {21}},\
  \bibinfo {pages} {1801200} (\bibinfo {year} {2019})}\BibitemShut {NoStop}%
\bibitem [{\citenamefont {Rodriguez}\ \emph {et~al.}(2025)\citenamefont
  {Rodriguez}, \citenamefont {Gontard}, \citenamefont {Ma}, \citenamefont
  {Xu},\ and\ \citenamefont {Persson}}]{rodriguez2025determine}%
  \BibitemOpen
  \bibfield  {author} {\bibinfo {author} {\bibfnamefont {N.}~\bibnamefont
  {Rodriguez}}, \bibinfo {author} {\bibfnamefont {L.}~\bibnamefont {Gontard}},
  \bibinfo {author} {\bibfnamefont {C.}~\bibnamefont {Ma}}, \bibinfo {author}
  {\bibfnamefont {R.}~\bibnamefont {Xu}}, \ and\ \bibinfo {author}
  {\bibfnamefont {B.}~\bibnamefont {Persson}},\ }\href@noop {} {\bibfield
  {journal} {\bibinfo  {journal} {Tribology Letters}\ }\textbf {\bibinfo
  {volume} {73}},\ \bibinfo {pages} {18} (\bibinfo {year} {2025})}\BibitemShut
  {NoStop}%
\bibitem [{\citenamefont {Gadelmawla}\ \emph {et~al.}(2002)\citenamefont
  {Gadelmawla}, \citenamefont {Koura}, \citenamefont {Maksoud}, \citenamefont
  {Elewa},\ and\ \citenamefont {Soliman}}]{gadelmawla2002roughness}%
  \BibitemOpen
  \bibfield  {author} {\bibinfo {author} {\bibfnamefont {E.~S.}\ \bibnamefont
  {Gadelmawla}}, \bibinfo {author} {\bibfnamefont {M.~M.}\ \bibnamefont
  {Koura}}, \bibinfo {author} {\bibfnamefont {T.~M.}\ \bibnamefont {Maksoud}},
  \bibinfo {author} {\bibfnamefont {I.~M.}\ \bibnamefont {Elewa}}, \ and\
  \bibinfo {author} {\bibfnamefont {H.~H.}\ \bibnamefont {Soliman}},\
  }\href@noop {} {\bibfield  {journal} {\bibinfo  {journal} {Journal of
  materials processing Technology}\ }\textbf {\bibinfo {volume} {123}},\
  \bibinfo {pages} {133} (\bibinfo {year} {2002})}\BibitemShut {NoStop}%
\bibitem [{\citenamefont {Brown}\ and\ \citenamefont
  {Cox}(2009)}]{c22_brown2009innovative}%
  \BibitemOpen
  \bibfield  {author} {\bibinfo {author} {\bibfnamefont {D.}~\bibnamefont
  {Brown}}\ and\ \bibinfo {author} {\bibfnamefont {A.~J.}\ \bibnamefont
  {Cox}},\ }\href@noop {} {\bibfield  {journal} {\bibinfo  {journal} {The
  Physics Teacher}\ }\textbf {\bibinfo {volume} {47}},\ \bibinfo {pages} {145}
  (\bibinfo {year} {2009})}\BibitemShut {NoStop}%
\bibitem [{\citenamefont {Moraru}\ and\ \citenamefont
  {Moraru}(2021)}]{c23_moraru2021distance}%
  \BibitemOpen
  \bibfield  {author} {\bibinfo {author} {\bibfnamefont {P.-G.}\ \bibnamefont
  {Moraru}}\ and\ \bibinfo {author} {\bibfnamefont {P.}~\bibnamefont
  {Moraru}},\ }\href@noop {} {\bibfield  {journal} {\bibinfo  {journal}
  {Issues’ 21-Issues in Education}\ ,\ \bibinfo {pages} {119}} (\bibinfo
  {year} {2021})}\BibitemShut {NoStop}%
\bibitem [{\citenamefont {Rodrigues}\ and\ \citenamefont
  {Carvalho}(2013)}]{c24_rodrigues2013teaching}%
  \BibitemOpen
  \bibfield  {author} {\bibinfo {author} {\bibfnamefont {M.}~\bibnamefont
  {Rodrigues}}\ and\ \bibinfo {author} {\bibfnamefont {P.~S.}\ \bibnamefont
  {Carvalho}},\ }\href@noop {} {\bibfield  {journal} {\bibinfo  {journal}
  {Physics Education}\ }\textbf {\bibinfo {volume} {48}},\ \bibinfo {pages}
  {431} (\bibinfo {year} {2013})}\BibitemShut {NoStop}%
\bibitem [{\citenamefont {Wee}\ \emph {et~al.}(2012)\citenamefont {Wee},
  \citenamefont {Chew}, \citenamefont {Goh}, \citenamefont {Tan},\ and\
  \citenamefont {Lee}}]{c25_wee2012using}%
  \BibitemOpen
  \bibfield  {author} {\bibinfo {author} {\bibfnamefont {L.~K.}\ \bibnamefont
  {Wee}}, \bibinfo {author} {\bibfnamefont {C.}~\bibnamefont {Chew}}, \bibinfo
  {author} {\bibfnamefont {G.~H.}\ \bibnamefont {Goh}}, \bibinfo {author}
  {\bibfnamefont {S.}~\bibnamefont {Tan}}, \ and\ \bibinfo {author}
  {\bibfnamefont {T.~L.}\ \bibnamefont {Lee}},\ }\href@noop {} {\bibfield
  {journal} {\bibinfo  {journal} {Physics Education}\ }\textbf {\bibinfo
  {volume} {47}},\ \bibinfo {pages} {448} (\bibinfo {year} {2012})}\BibitemShut
  {NoStop}%
\end{thebibliography}%
\bibliographystyle{apsrev4-1}
\end{document}